\documentclass[showpacs,onecolumn,aps,floatfix,superscriptaddress,noshowpacs]{revtex4}
\usepackage[mathlines]{lineno}
\usepackage{amsthm}
\usepackage{soul}
\usepackage{dsfont}
\usepackage{amsmath,amssymb,graphicx,bm,color,mathrsfs,verbatim,epstopdf,dcolumn,cancel}
\usepackage{bbold}
\usepackage{multirow}
\usepackage{pgfmath}
\usepackage{ytableau}
\ytableausetup{centertableaux}

\usepackage{graphicx}
\usepackage{tikz-cd}

\DeclareMathOperator{\Hom}{Hom}

\DeclareMathOperator{\Res}{Res}

\usepackage[fleqn]{mathtools}

\newtheorem{theorem}{Theorem}

\usepackage{ulem,cancel,comment}

\graphicspath{ }

\usepackage{hyperref}
\hypersetup{ 
	colorlinks   =  true
}

\usepackage[fleqn]{mathtools}

\begin{document}

\title{ $k$-Fold Gaussian Random Matrix Ensembles I: \\ Forcing Structure into Random Matrices }

\author{Michael Brodskiy}
\affiliation{College of Engineering, Northeastern University, 360 Huntington Ave., Boston, MA 02115, USA}
\author{Owen L. Howell}
\affiliation{Department of Electrical and Computer Engineering, Northeastern University, 360 Huntington Ave., Boston, MA 02115, USA}
\begin{abstract}
Random Matrix Theory is a powerful tool in applied mathematics. Three canonical models of random matrix distributions are the Gaussian Orthogonal, Unitary and Symplectic Ensembles. For matrix ensembles defined on $k$-fold tensor products of identical vector spaces we motivate natural generalizations of the Gaussian Ensemble family. We show how the $k$-fold invariant constraints are satisfied in both disordered spin models and systems with gauge symmetries, specifically quantum double models. We use Schur-Weyl duality to completely characterize the form of allowed probability distributions. The eigenvalue distribution of our proposed ensembles is computed exactly using the Harish-Chandra integral method. For the $2$-fold tensor product case, we show that the derived distribution couples eigenvalue spectrum to entanglement spectrum. Guided by representation theory, our work is a natural extension of the standard Gaussian random matrix ensembles.
\end{abstract}
\date{\today}
\maketitle

\section{Introduction}

Random matrices were originally introduced by Eugene Wigner in the 1950s to study the spacing of energy levels of heavy atomic nuclei \cite{Wigner_1955}. Random matrices have long since outgrown nuclear physics and Wigner's random matrix ensembles have found uses in a diverse set of fields, from pure mathematics \cite{Berry_2023_Riemann,Rudnick1996ZerosOP,Sierra_2011,Berry1999Hequalsxp,Conrey2006Random,Keating2006number} to the physics of materials \cite{Efetov_2005,Pan_2022,Byrnes_2022,Feinberg_2021,Kanzieper_1998,Stotland_2008,Comets_2019,Beenakker1997Randomquantumtransport,Guhr1998RMTquantum,Altland1997Nonstandard}. Random matrices have found applications far removed from the physical sciences: eigenvalue repulsion is displayed in bus waiting times \cite{Krblek2000mexico}, and random matrices play a role in both financial modeling \cite{bouchaud2009financial,potters2005financial} and network theory \cite{Kuhn2011Spectra,Ergun2009spectra,van_Handel2017Structured,Newman2006Modularity,hartwell1999molecular}. Symmetry serves as a guiding principle in development of random matrix theory, and random matrix ensembles are almost completely characterized by what set of symmetries they are invariant under. The Dyson `3-fold' way \cite{Dyson_2004_Threefold} gives a classification of random matrix ensembles with distributions invariant under unitary, orthogonal or symplectic conjugation. Similarly, \cite{Altland1997Nonstandard} proposed the famous `10-fold' way which characterizes random matrix ensembles under time-reversal, parity and charge conjugation. In this note, we consider a natural generalization of \cite{Dyson_2004_Threefold} which considers random matrix ensembles invariant under local, as opposed to global, symmetry transformations.

\subsection{Random Matrix Theory in Pure Mathematics}
Random matrix theory also serves as an important aspect of pure mathematics. Specifically, many of the formal results of `universality' are understood through the lens of random matrix theory \cite{tao2021}. The Wiengarten calculus is a systemic computational method for evaluating integrals over the unitary group \cite{Collins_2016}. \cite{Vleeshouwers_2023} shows how representation theoretic methods can be used to evaluate integrals over unitary group to symmetric polynomials. In a similar manner, \cite{Collins_2023,Collins_2023_II} considered a generalization of the Harish-Chandra integral formula to the tensor product case. The integrals in \cite{Collins_2023_II,Collins_2023} are similar to those we consider in this note.

\subsection{Random Matrix Theory in Quantum Information}
Recently, there has been extensive theoretical work in quantum information theory studying random unitary evolution \cite{Collins_2016,Chen_2022,Li_2022,Mitchell_2005,Gharibyan_2018}. Randomly drawn unitary matrices have been used to model the evolution of generic quantum systems. A recent breakthrough in the development of a theory of quantum chaos was the development of the entanglement membrane description \cite{Zhou_2020Entanglment,Gong_2022}, which is an effective theory of quantum chaos. The entanglement tension has been calculated for random unitary evolution \cite{Zhou_2020Entanglment,Sierant2023Entanglement}, as well as integrable models \cite{Rampp2023Entanglement}. A very interesting research direction is the holographic description of the entanglement membrane theory \cite{Blake2023Page,Mezei_2018,Ag_n_2021}. Much of the interest in these random unitary evolutions has been spurred by the study of chaos in thermofield double state \cite{Cottrell_2019theromfield,Shenker2014Black,Aalsma2021Shocks}. The evolution of thermofield double states serves as a probe of quantum chaos \cite{Shenker2014Black,Aalsma2021Shocks}. Thermofield double states also serve as a dual theory to traversable wormhole \cite{Gao_2017_Traversable,Gao2021Traversable,Xu2020Sparse}. Signatures of synthetic traversable wormholes have been observed experimentally \cite{Jafferis_2022_Traversable}.

Furthermore, gravitational models serve as dual theories for many natural random matrix models \cite{stanford2020jt,Turiaci2023n2JTsupergravity}. Random matrix theories can be formulated as (0+0)-dimensional field theories \cite{Zee_2016}. Field theory techniques in random matrix theory have been utilized as a powerful tool for studying thermilization properties of quantum systems \cite{Jafferis2023jt_gravity,Jafferis2023MatrixModels}. Specifically, as first observed in \cite{saad2019jt} many random matrix models can be realized as theories dual to Jackiw–Teitelboim gravity \cite{Mann1989gh,Jafferis2023MatrixModels,Turiaci2023n2JTsupergravity}. We comment on the diagrammatic rules for our proposed ensembles in Section \ref{Field_Theory_Description}. Our results describe a natural new class of non-interacting field theories for random matrix models.



\section{Symmetry and Invariance in Random Matrix Theory}

Random matrix ensembles are almost completely defined by the class of transforms they are invariant under. Specifically, the Gaussian Ensembles are the unique non-commutative probability distribution with independent matrix elements that is invariant under normal transformations \cite{Porter1960NuclearSpectra}. For this reason, group theory plays a central role in random matrix theory. The three canonical Gaussian ensembles, are characterized by their invariance of measure under the orthogonal, sympletic and unitary groups \cite{Dyson_2004_Threefold}. Group and representation theoretic tools have similarly been used to study random matrices. Many of the representation theoretic tools that we use to derive our results have been previously applied to study quantum systems. Schur-Weyl methods have been used to understand thermodynamic properties of quantum systems \cite{Yadin_2023Thermodynamics}. Similarly, the Harish-Chandra integral method was used to compute the eigenvalue-eigenvalue correlations in a model of coupled random matrices \cite{Itzykson1979Planar}. By enforcing a set of natural isotropy conditions, we derive a new class of random matrix ensembles that exhibit properties of local quantum systems. The constraints we impose are less restrictive than the Gaussian Ensemble constraints, which allows for terms not allowed in standard Gaussian ensembles. For this reason, our proposed ensembles exhibit features not seen in the Gaussian ensemble family.

\subsubsection{ Random Matrices versus Real Quantum Systems }
In many ways, the Gaussian ensembles are too `coarse' to describe physical systems of interest in condensed matter, as many of the features that we are interested in, such as approximate local integrals of motion, are not captured by the Gaussian ensembles. Specifically, the GUE and GOE ensembles do not contain information about locality, and do not capture low-energy properties observed in quantum systems. As an example of the discrepancy between GUE and real quantum systems, the Berry conjecture \cite{Stockmann1999Quantum,Srednicki_1994Chaos} essentially states that the high energy eigenfunctions behave as if they were Gaussian random variables. This property is emphatically \emph{not} observed in existing random matrix models, where there are no correlations between eigenvalue and eigenvector structure. Specifically, the only terms allowed under general unitary invariance depends only on sums powers of eigenvalues. In section \ref{Eigenvalue-Entanglement Interactions}, we show that our proposed ensembles allow for terms that directly couple entanglement related quantities to eigenvalues, forcing large spectrum eigenvectors to have random coefficients.

\begin{figure}[h]
	\centering
	\begin{tabular}{ccc}
		\includegraphics[width=0.33\textwidth]{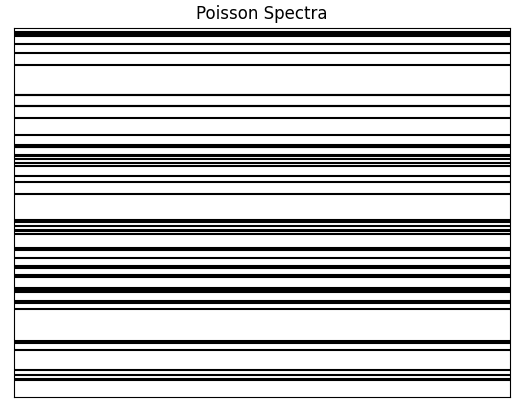}  &
		\includegraphics[width=0.33\textwidth]{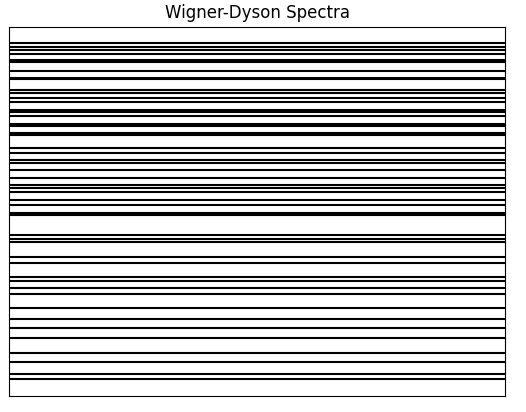} 
		\includegraphics[width=0.33\textwidth]{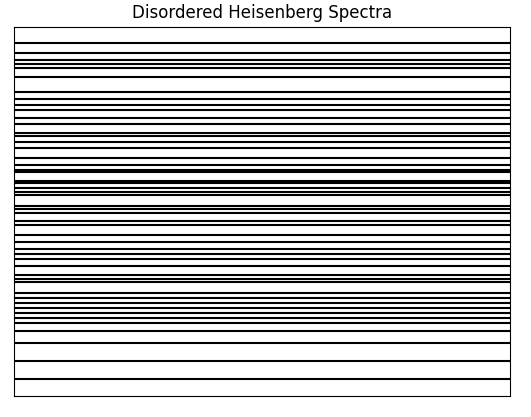} 
	\end{tabular}
	\caption{ Left: Spectrum with Poisson Level Spacing. Center: Spectra with Wigner-Dyson Level-Spacing. Right: Spectra of Disordered Heisenberg Model (See \ref{Disordered_Heisenberg} for definition ). The Disordered Heisenberg model has a broken $k$-fold $SO(3)$-symmetry, leading to approximate integrability in the low-energy sector of the spectrum. Note that the low-energy spectra of the disordered Heisenberg model has much more regular eigenvalue spacing than the Wigner-Dyson distribution. This is because the Heisenberg model has an approximate $SO(d)$-symmetry.   }
	\label{fig:Symmetrized_Errors}
\end{figure}

\begin{table}
	\centering
	\begin{tabular}{ |p{4.23cm}|c|p{4.0cm}|p{3.5cm}|  }
		\hline
		\multicolumn{4}{|c|}{ \textbf{Comparison of Assumptions in Random Unitary Matrix Models on $W = V_{L} \otimes V_{R} $}  } \\
		\hline
		Ensemble Name & Elementwise Independence & Invariant Transformation & Subspace correlations   \\
		\hline
		\hline
		$ \text{GUE}(V_{L}) \otimes  \text{GUE}(V_{R}) $ & Yes  & $H \rightarrow ( U_{L} \otimes U_{R} ) H ( U_{L} \otimes U_{R} )^{\dagger} $& Completely Independent   \\
		\hline
		$\text{GUE}( V_{L} \otimes V_{R} )$ & Yes  & $H \rightarrow UHU^{\dagger}$  & Indepdendent    \\ 
		\hline
		\hline
		\multicolumn{4}{|c|}{ \textbf{Special Case of Isomorphic Subspaces: $ V_{R}\cong V_{L} $ } } \\
		\hline
		\hline
		$\text{2-Fold GUE}( V_{L} \otimes V_{R} )$ & No  &  $H \rightarrow ( V \otimes V ) H ( V \otimes V )^{\dagger} $ &   Non-Trivial   \\
		\hline
		$\text{twisted 2-Fold GOE}( V_{L} \otimes V_{R} )$ & No  &  $H \rightarrow (V \otimes V^{\dagger} ) H (V \otimes V^{\dagger} )^{\dagger} $ &   Non-Trivial   \\
		\hline
	\end{tabular}
	\caption{  Comparison of properties of random matrix models defined on the $W = V_{L} \otimes V_{R}$ vector space. The matrices $U , U_{L} , U_{R}$ are unitary matrices. The unitary matrix $U$ is defined on $W = V_{L} \otimes V_{R}$. The unitary matrix $U_{R}$ is defined on the vector space $V_{R}$. The unitary matrix $U_{L}$ is defined on the vector space $V_{L}$. In the special case where $V_{L} \cong V_{R}$ are isomorphic vector spaces, there are additional natural ensembles based on invariance under the tensor product representation. The unitary matrix $V$ is defined on the $V_{L} \cong V_{R}$ vector space. }
\end{table}

\subsubsection{ Random Matrices in Learning Theory }

Outside of the physical sciences, the theory of random matrices has found extensive use within the statistical learning community. Specifically, GOE ensembles arise naturally within the context of error distributions for a variety of recovery problems. Random matrices have also found usage in deep learning as models for weight matrices of deep neural networks \cite{Thamm_2022,Dohmatob_2022,Advani_2017,Louart_2017}. Random matrices have also been used to study the asymptotic risk of transfer learning techniques \cite{Yang_2020}.The spectra of Hessian matrices of deep neural networks are well described by the Gaussian orthogonal ensemble \cite{baskerville2023random}. There have been suggestions that genrelizability of neural networks can be diagnosed via eigenvalue distributions of neural network weights \cite{Wei_2022,Martin_2021_Implict}. Advancements in network initialization to ensure non-vanishing and non-exploding gradients come from the assumption that weights are initialized as random Gaussian matrices \cite{Baskerville_2022,Baskerville_2022_2}. A interesting line of research parameterizes the output of a neural network as a contraction of tensors living on a $V^{\otimes k}$ tensor product space \cite{Alexander2023Makes}. In this work, we motivate and propose a new class of random matrices that are generalizations of the Gaussian Ensemble models that are invariant under $k$-fold normal transformations. We conjecture that these ensembles will find use in situations with inherent tensor product structures.

\begin{figure}[h]
\centering
\fbox{ \includegraphics[width=1.0\textwidth]{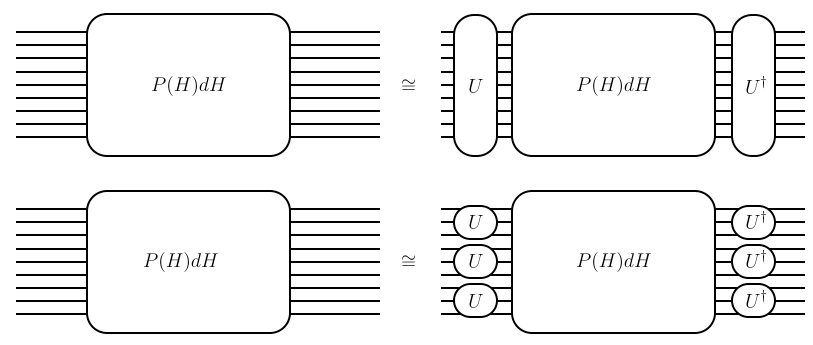} }
\caption{Tensor Diagrammatic of Proposed Matrix Ensembles. Random matrix ensembles are almost completely defined by the class of transforms they are invariant under. Top: The standard Gaussian Ensembles are invariant under generic change of basis $H \rightarrow UHU^{\dagger}$. Bottom: We can define new random matrix ensembles by requiring invariance under the local change of basis $H \rightarrow U^{\otimes k} H ( U^{\otimes k} )^{\dagger} $ (in this case $k=3$ and local vector space dimension $d=3$, so that $U$ is a $3\times3$ unitary matrix). }
\label{fig:Diagrammatic}
\end{figure}

\section{ Tensor Product Vector Spaces}

Let $V$ be a vector space over $\mathbb{R}$ or $\mathbb{C}$. In many linear algebra applications, we often work with a vector space $W$ that is composed of $k$-fold tensor products of the smaller vector space $V$ such that
\begin{align*}
	W = V^{\otimes k} = \underbrace{ V \otimes V \otimes ... \otimes V }_{ k-\text{times} }
\end{align*}
where the dimension of the vector space $V$ is $d$, $\dim V = d$ so that the dimension of $W$ is $\dim W = d^{k}$. This situation arises naturally in dealing with quantum mechanical systems of many identical particles.

\subsection{Random Matrix Distributions on $ V^{\otimes k} = V \otimes V ... \otimes V$ Vector Spaces}

Let $V$ be a vector space defined over the field $\mathbb{R}$ or $\mathbb{C}$. Let $W = V^{\otimes k}$. The standard Gaussian Ensembles ignore the tensor product structure of the underlying vector space. In many ways, these ensembles violate the principle of maximum entropy \cite{Jaynes_1982} as the information that $W = V^{\otimes k} = V \otimes V \otimes ... \otimes V $ is discarded. We would like to define a new class of random matrix ensembles that respect the underlying structure of the $W = V^{\otimes k}$ vector space.

\noindent
One option is to independently draw $k$ random matrices from a Gaussian ensemble on $V$ and form their tensor product. This method is rather naive, as there are no correlations between matrices on different $V$ subspaces. Another possible approach to introduce a matrix ensemble with correlations between the different $V$ subspaces would be to draw a matrix $H$ from a canonical distribution on $V$ and form the $k$-fold tensor product $H^{\otimes k} = H \otimes H \otimes ... \otimes H $. However, this proposed distribution is inadequate as the eigenvalue spectrum of $H^{\otimes k}$ is just the $k$-fold spectrum of $H$ ( and other than this new $k$-fold degeneracy ) no new features arise that are not present in the standard Gaussian ensembles. In this note, we propose a new class of random matrix distributions that naturally interpolates between these two extremes: The distribution we propose is invariant under $k$-fold normal transformations of the form
\begin{align}\label{k_fold_normal_transform}
	\forall U \in G, \enspace	X \rightarrow U^{\otimes k} X (U^{\otimes k})^{\dagger} =  \underbrace{(U \otimes U \otimes ... \otimes U)}_{k-\text{times}} X \underbrace{(U \otimes U \otimes ... \otimes U)^{\dagger}}_{k-\text{times}}
\end{align}
where $U^{\otimes k} = (U \otimes U \otimes ... \otimes U) $ denotes the $k$-fold tensor product. The group $G$ is chosen to be either the unitary group $U(d)$, the orthogonal group $O(d)$ or the symplectic group $Sp(d)$, although in principal $G$ can be any group. When $G$ is unitary, orthogonal or sympletic, we will define $H$ on Hermitian, symmetric or skew-symmetric matrices, respectively.

The constraint \ref{k_fold_normal_transform} can alternatively be viewed as a non-commutative probability distribution that is invariant under a local $G$ degree of freedom (this is explained in more depth in \ref{Appendix:Physical_Models} ). Standard random matrix ensembles have no concept of locality, but by restricting demanding that our ensemble be invariant under \ref{k_fold_normal_transform} instead of the standard unitary conjugation, we can see the existence of gappless modes, as predicted by Goldstone's theorem \cite{Goldstone1962Broken,Zee_2016}. Specifically, invariance under \ref{k_fold_normal_transform} implies that a redefinition of each local Hilbert space in the same way leaves the matrix ensemble density unchanged. This is explained in more depth with multiple concrete physical models in \ref{Appendix:Physical_Models}. We summarize our contributions as follows:
\begin{itemize}
	\item Inspired by recent work on free probability and random matrix theory, \cite{Jafferis2023MatrixModels,Pappalardi_2022_Eigenstate}, we propose a new class of random matrix models that are invariant under $k$-fold normal transformation. We show that many physical models of interest satisfy our desired properties.
	\item Using Schur-Weyl duality, we give a complete characterization of matrix models which satisfy our desiderata. We show that under a reasonable set of assumptions, our construction is unique.
	\item Using a generalization of the Harish-Chandra method, we derive the eigenvalue distribution of our proposed ensembles. We compare the derived eigenvalue distribution with the standard Wigner-Dyson statistics \cite{Mehta_2004} and the $(\alpha,\beta)$-ensembles \cite{Altland1997Nonstandard}.
\end{itemize}

\section{Matrix Invariant Theory}

We will be interested in characterizing random matrix distributions that are invariant under the $k$-fold normal transformation \ref{k_fold_normal_transform} for the unitary, orthogonal and symplectic groups. For the standard $k=1$ case, a theorem of Weyl gives a complete characterization of matrix valued functions invariant under matrix conjugation,
\begin{theorem}[ Invariant Matrix Polynomials \cite{Weyl_1966_Classical} ]\label{Weyl_Lemma}
	All the invariants of an ($d\times d$)-matrix $H$ under the non-singular similarity transformation of $U$,
	\begin{equation}
		\forall U \in GL(d), \quad    H \rightarrow H' = U H U^{-1}
	\end{equation}
	can be expressed as functions of the traces of the first $d$-powers of the matrix $H$.
\end{theorem}
\noindent
This theorem gives characterization of matrix invariant quantities under $GL(d)$ transformation. Matrix invariants of subgroups of $GL(d)$ can be computed via the Chevalley restriction theorem \cite{Weyl_1966_Classical,Hall_2015_Lie}.
\begin{theorem}[ Chevalley Restriction Theorem  (cite) ]\label{Chevalley_Lemma}
	Let $G$ be a compact Lie group. Let $W(G) \subseteq G$ be the Weyl group of $G$. Let $\frak{g}$ be the Lie algebra of $G$. Let $\frak{h}$ be a Cartan subalgebra of $\frak{g}$. Then, there is an isomorphism of invariants
	\begin{equation}
		\mathbb{C}[ \mathfrak{g} ]^{G} \cong \mathbb{C}[ \mathfrak{h} ]^{W(G)}
	\end{equation}
	so that the space invariant group algebra elements of $G$ is isomorphic to the space of invariant group algebra elements of the Weyl group $W(G)$.
\end{theorem}
\noindent
The Chevalley theorem \ref{Chevalley_Lemma} allows for calculation of $G$-invariants in terms of $W(G)$-invariants. Because the Weyl group $W(G)$ is abelian, this is a much simpler calculation.

\subsection{Characterization Theorems}
Theorems \ref{Weyl_Lemma} and \ref{Chevalley_Lemma} restricts the form of allowed random matrix distributions invariant under unitary transformation. When the assumption of element-wise independence is made, the probability distribution takes the form
\begin{align*}
	\text{Wigner Distribution: } \text{Pr}[H] dH \propto \exp( - \lambda \text{Tr}[ H^{2} ] ) dH
\end{align*}
for some $\lambda >0 $. A result from \cite{Porter1960NuclearSpectra} states that the Wigner distribution is the \emph{unique} distribution with element-wise independence that is invariant under unitary change of basis.

\begin{theorem}[ Uniqueness of Wigner Distribution \cite{Porter1960NuclearSpectra}  ]\label{Uniqueness}
	Let $P(H)dH$ be a probability measure on random Hermitian/Symmetric/Skew-Symmetric matrices. Suppose that the measure satisfies the two properties: \\
	I.The probability measure is invariant under the unitary/orthogonal/symplectic change of basis, 
	\begin{align*}
		H' = UHU^{\dagger} , \quad P(H')dH' = P(H)dH
	\end{align*}
	so that the measure is invariant under conjugation by all unitary/orthogonal/symplectic matrices $U$. \\
	II. The matrix elements $H_{ij} = \bar{H}_{ji}$/$H_{ij} = H_{ji}$/$H_{ij} = -H_{ji}$ are statistically independent and Gaussian distributed. \\
	If both properties I-II are satisfied then the $P(H)dH$ is the Gaussian Unitary/Orthogonal/symplectic Ensemble.
\end{theorem}

\noindent
We would like a result that generalizes \ref{Uniqueness} to the weaker constraint \ref{k_fold_normal_transform}. We prove a analogy of theorem \ref{Weyl_Lemma} for $k$-fold tensor spaces, 
\begin{theorem}[ $k$-Fold Invariant Matrix Polynomials  ]\label{k-fold_Weyl_Lemma}
	For each permutation $\sigma \in S_{k}$, define the permutation operators $\hat{S}_{\sigma}$ as the operator that has action
	\begin{align*}
		\hat{S}_{\sigma} | i_{1} , i_{2} , ... , i_{k} \rangle = | i_{\sigma(1)} , i_{\sigma(2)} , ... , i_{\sigma(k)} \rangle
	\end{align*}
	Then, all the invariants of an ($d^{k} \times d^{k}$)-matrix $H$ under the non-singular $k$-fold similarity transformation of $U$,
	\begin{equation}
		\forall U \in GL(d), \quad    H \rightarrow H' = U^{\otimes k} H ( U^{\otimes k} )^{-1}
	\end{equation}
	can be expressed as functions of the traces of the first $d$-powers of the set of matrices $\{ H_{\sigma} = S_{\sigma}HS^{\dagger}_{\sigma} \enspace | \enspace \sigma \in S_{n} \}$.
\end{theorem}
This result is derived via the Schur-Weyl lemma, which is a generalization of \ref{Weyl_Lemma} to tensor products of non-singular matrices. 
\begin{theorem}[Schur-Weyl Lemma \cite{Roberts_2017_Chaos} ]\label{Schur-Weyl_Lemma}
	Let $X$ be a matrix that commutes with the tensor product of $k$-fold tensor products of non-singular matrices,
	\begin{align*}
		\forall U \in GL(d), \quad U^{\otimes k}X = X U^{\otimes k}
	\end{align*}
	then, the matrix $X$ can be written as a linear sum of permutation operators $X = \sum_{ \sigma \in S_{k} } c_{\sigma} \hat{S}_{\sigma}$
\end{theorem}
\noindent
The coefficients $c_{\sigma}$ in the expansion of $X$ can be computed in terms of the form $\text{Tr}[ \hat{S}_{\sigma} X ]$ \cite{Pappalardi_2022_Eigenstate} (see also \ref{More_Schur-Weyl_Duality} for a review of Schur-Weyl duality ).

\section{ $k$-Fold Gaussian Ensembles}
The goal of this section is to establish a desiderata that we would like our proposed models to have. We define the $k$-Fold Gaussian Ensemble as the matrix distribution satisfying the following four properties:
\begin{itemize}
	\item Property I: Invariance of Measure under $k$-fold Conjugation
	\item Property II: Gaussian Distribution of Matrix Elements
	\item Property III: Subspace Homogeneity
\end{itemize}

\noindent
\subsection*{Property I: Invariance of Measure under $k$-fold Conjugation}\label{Property I}
\noindent
Let $G$ be either the unitary, orthogonal or symplectic group. Let $P(H)dH$ be a probability distribution on Hermitian matrices.  Consider the $k$-fold normal transformation,
\begin{align*}
	\forall U \in G, \quad H \rightarrow H' = U^{\otimes k} H ( U^{\otimes k } )^{\dagger}
\end{align*}
Under this transformation, the probability density satisfies,
\begin{align*}
	P(H)dH = P(H')dH'
\end{align*}
So that the probability measure is invariant under $k$-fold normal transformation. In \ref{Appendix:Physical_Models}, we give examples of models that satisfy this condition.

\subsection*{Property II: Gaussian Distribution of Matrix Elements }\label{Property II}
\noindent
Let $H_{ij}$ with $i\leq j$ be the independent matrix elements of the Hermitian matrix $H = H^{\dagger}$. Let $\Vec{H} = \text{vec}(H)$ be the vectorization of each independent matrix element of $H$. Then, the probability density can be written as a multivariate Gaussian distribution,
\begin{align*}
	P(H)dH \propto \exp( - \frac{1}{2} \Vec{H}^{\dagger} \Delta \Vec{H} ) \prod_{i\leq j} dH_{ij}
\end{align*}
where the matrix $\Delta$ is Hermitian $\Delta = \Delta^{\dagger}$ and positive definite $\Delta \succ 0$. For notational convenience we have suppressed the tensor product indices $i=i_{1}i_{2}...i_{k}$ and $j=j_{1}j_{2}...j_{k}$ so that
\begin{align*}
	\prod_{i\leq j} dH_{ij} = \prod_{i_{1}\leq j_{1}}\prod_{i_{2}\leq j_{2}}...\prod_{i_{k}\leq j_{k}} dH_{i_{1}i_{2}...i_{k}j_{1}j_{2}...j_{k}}
\end{align*}

\subsection*{Property III: Subspace Homogeneity }\label{Property III}
\noindent
We demand that the distribution is invariant under permutation of subspaces. Specifically, each tensor product subspace should be identical. Under a permutation of subspaces,
\begin{align}\label{Homogeneity}
	\forall \sigma \in S_{k}, \quad H \rightarrow H' = S_{\sigma} H S^{\dagger}_{\sigma}
\end{align}
we require that the probability density is invariant $P(H)dH = P(H')dH'$. This constraint places a additional restriction on the allowed form of the probability density. The standard Gaussian ensembles satisfy this property. This property was noted in \cite{Yadin_2023Thermodynamics}, which noted that although observable quantities need to be permutation invariant, it is possible that underlying states have emergent exotic symmetry.

\subsubsection{Alternate Swap Constraints?}
For the unitary and sympletic ensembles, there is an alternative natural constraint to \ref{Homogeneity}. Let $\text{Sign} : S_{k} \rightarrow \pm 1$ be the sign representation of the symmetric group of order $k$. 
\begin{align}\label{Sign_Homogeneity}
	\forall \sigma \in S_{k}, \quad H \rightarrow \text{Sign}(\sigma) S_{\sigma} H S^{\dagger}_{\sigma}
\end{align}
where the distribution is odd under permutation of two subspaces. Bosonic and fermionic statistics correspond to the trivial and sign representations of $S_{n}$ and \ref{Homogeneity} and \ref{Sign_Homogeneity} can be thought of as bosonic and fermionic exchanges, respectively. As an aside, it may be possible to generalize \ref{Homogeneity} and \ref{Sign_Homogeneity} to include non-abelian statistics, using higher dimensional representations of $S_{n}$. Usually topological quantum computing models are formulated in terms of unitary evolutions. For this reason, it makes more sense to formulate alternative statistics using Dyson's circular ensembles \cite{Dyson_2004_Threefold,Mehta_2004} although we will leave this direction for future work.

\subsubsection{Interpretation of Desiderata}

We comment on the interpretation of properties I-IV. Property \ref{Property I} is the natural generalization of the standard change of basis invariance property, which is motivated in section I. Property \ref{Property II} demands that individual matrix elements are a (possibly correlated) Gaussian distribution. \ref{Property III} requires that the probability density is independent of the labeling of the tensor product subspaces. 


\subsection{Characterization Theorem}

We give a characterization of random matrix distributions that satisfy our desired properties. We state theorem \ref{Charecterization_Lemma}, which we derive in the next section.

\begin{theorem}[ Characterization of $k$-fold Gaussian Ensembles   ]\label{Charecterization_Lemma} Let $G$ be the unitary/orthogonal group/symplectic. All matrix probability distributions satisfying properties I-IV can be written uniquely in the form
	\begin{align}\label{probility_density}
		P(H)dH \propto \exp( - \frac{1}{2} \Vec{H}^{\dagger} \Delta \Vec{H}  )dH
	\end{align}
	where the matrix $\Delta$ takes the form
	\begin{align}\label{Covarience_Matrix}
		\Delta = U[  \bigoplus_{ \mu \mu' } \bigoplus_{s \in \pm} M_{\mu \mu' s} \otimes \mathbb{1}_{d_{\mu}} \otimes \mathbb{1}_{ d'_{\mu'} }   ]U^{\dagger}
	\end{align}
	where $U$ is a fixed unitary/orthogonal/skew-orthogonal matrix and $M_{\mu \mu' s}$ are positive definite Hermitian/symmetric/skew-symmetric random matrices of dimension $C_{\mu \mu' s} \times C_{\mu \mu' s}$ where $C_{\mu \mu' s}$ is completely determined by representation theory of the group $G$. $d_{\mu} = \dim V_{\mu}$ and $d'_{\mu'} = \dim \mu'$ are the dimensions of the irreducible $G$ and $S_{k}$ representations, respectively.
\end{theorem}
\noindent 
Probability densities of the form \ref{probility_density} are Gaussian distributed and the all moments can be computed exactly. Specifically, let $\vec{J} = \text{Vec}( J )$. The generating function
\begin{align*}
	Z( \Vec{J} ) = \int dH \exp( - \frac{1}{2} \Vec{H}^{\dagger} \Delta \Vec{H} + \Vec{J}^{\dagger} \Vec{H}  )
\end{align*}
has closed form solution given by
\begin{align*}
	Z(J) = \int dH \exp( - \frac{1}{2} \Vec{H}^{\dagger} \Delta \Vec{H} + \Vec{J}^{\dagger}\Vec{H}  ) = \frac{ (2\pi)^{ \frac{dk}{2} } }{ \det( \Delta )^{\frac{1}{2}} } \exp( \frac{1}{2} \Vec{J}^{\dagger} ( \Delta )^{-1} \Vec{J}  )
\end{align*}
The generating function $Z(J)$ specifies all correlation functions of the theory. When interaction terms are added, $Z(J)$ is used to derive the Feynman rules. If we change basis to $\Vec{J}' = U \Vec{J}$ then we have that
\begin{align*}
	Z( \Vec{J}' ) = \frac{ (2\pi)^{ \frac{dk}{2} } }{ \prod_{\mu \mu' s} \det( M_{\mu \mu' s} )^{\frac{1}{2}} } \exp( \frac{1}{2} ( \Vec{J}' )^{\dagger} [ \bigoplus_{\mu \mu' s} M^{-1}_{\mu \mu' s} \otimes \mathbb{1}_{d_{\mu}} \otimes \mathbb{1}_{d'_{\mu'}}  ] \Vec{J}'  )
\end{align*}
So that the generating function of $\Vec{J}'$ breaks down into a product
\begin{align*}
	Z( \Vec{J}' ) = \prod_{ \mu \in \hat{G} } \prod_{ \mu' \in \hat{S}_{k} } \prod_{s \in \pm } Z_{\mu \mu' s}( \Vec{J}'_{\mu \mu' s} )  
\end{align*}
where $\Vec{J}' = \bigoplus_{ \mu \mu' s } \Vec{J}'_{\mu \mu' s} $ decomposes into a direct sum of independent random variables. The generating function for $\Vec{J}'_{\mu \mu' s}$ is given by
\begin{align*}
	Z_{\mu \mu' s}( \Vec{J}'_{\mu \mu' s} ) = \frac{1}{ \det( M_{\mu \mu' s} )^{\frac{1}{2}} }  \exp( \frac{1}{2} ( \Vec{J}'_{\mu \mu' s} )^{\dagger} [ M^{-1}_{\mu \mu' s} \otimes \mathbb{1}_{d_{\mu}} \otimes \mathbb{1}_{d'_{\mu'}}  ] \Vec{J}'_{\mu \mu' s}  )
\end{align*}
Partition functions thus factorize into sectors labeled by irreducible representations of the unitary group $U(d)$. When $k=1$, this result reduces to the partition function of the standard Gaussian unitary ensemble.

\subsubsection{Invariance of Measure Under $k$-Fold Normal Transformation}
To begin, consider property I and property II. Under a $k$-fold normal transformation, we show that the measure $dH$ is invariant. Specifically, let 
\begin{align*}
	H' = U^{\otimes k} H ( U^{\otimes k} )^{\dagger}
\end{align*}
The matrix $U^{\otimes k}$ is a tensor product of unitary matrices and is itself unitary. Using a result of \cite{Mehta_2004}, the Jacobin of the transformation of any unitary conjugation is the identity, so $dH' = dH$. Similarly, under a permutation $H \rightarrow S_{\sigma}H S_{\sigma}^{\dagger}$ the measure is unchained as
\begin{align*}
	dH = \prod_{i \leq j} dH_{ij} \rightarrow \prod_{ i \leq j} dH'_{\sigma(i)\sigma(j)} = \prod_{i \leq j} dH_{ij} = dH'
\end{align*}
is just a re-ordering of indices and keeps the measure $dH$ invariant.

\subsubsection{Schur-Weyl Constraints on the Precision Matrix}

The joint requirements of $k$-fold invariance and subspace homogeneity place significant constraints on the allowed form of the covariance matrix $\Delta$. Specifically, the precision matrix $\Delta$ is required to satisfy the constraint
\begin{align*}
	\forall U \in G, \enspace \forall \sigma \in S_{k}, \quad [ U^{ \otimes 2k } , \Delta ] = 0 = [ \hat{S}_{\sigma} \otimes \hat{S}_{\sigma} , \Delta ] 
\end{align*}
Note that the $G$ action and the $S_{k}$ permutation action commute, \enspace
\begin{align*}
	\forall U \in G, \enspace \forall \sigma \in S_{k}, \quad [ U^{\otimes 2k} , \hat{S}_{\sigma} \otimes \hat{S}_{\sigma} ] = 0
\end{align*}
We can thus define the combined $G \times S_{k}$ group action on the space $w_{i_{1}i_{2}...i_{k}j_{1}j_{2}...j_{2k}} \in W = V^{\otimes 2k}$ as
\begin{align*}
	\forall U \in G, \enspace \forall \sigma \in S_{k}, \quad \Pi^{k}( U , \sigma) w_{i_{1}i_{2}...i_{k}j_{1}j_{2}...j_{2k}} = U_{i_{\sigma(1)}i'_{1}}U_{i_{\sigma(2)}i'_{2}}...U_{i_{\sigma(k)}i'_{k}}U_{j_{\sigma(1)}j'_{1}}U_{j_{\sigma(2)}j'_{2}}...U_{j_{\sigma(k)}j'_{k}}w_{i'_{1}i'_{2}...i'_{k}j'_{1}j'_{2}...j'_{2k}}
\end{align*}
There is also a $\mathbb{Z}_{2}$-action from swapping the $V^{\otimes k}$ subspaces. Specifically, define the $\hat{T}$ operator
\begin{align*}
	\forall w_{i_{1}i_{2}...i_{k}j_{1}j_{2}...j_{2k}} \in W = V^{\otimes 2k}, \quad \hat{T}w_{i_{1}i_{2}...i_{k}j_{1}j_{2}...j_{k}} = w_{j_{1}j_{2}...j_{k}i_{1}i_{2}...i_{k}}
\end{align*}
The operator $\hat{T}^{2} = \mathbb{I}$ is idempotent and so the set of operators $\{ \mathbb{I} , \hat{T} \}$ forms a representation of the group $\mathbb{Z}_{2}$. Note that $\hat{T}$ commutes with both the permutation and $k$-fold $G$ action. Specifically,
\begin{align*}
	\forall U \in G, \enspace \forall \sigma \in S_{k}, \quad [U^{\otimes 2k} , \hat{T} ] = 0 = [ \hat{S}_{\sigma} \otimes \hat{S}_{\sigma} , \hat{T} ] 
\end{align*}
Furthermore, the matrix $\Delta$ satisfies $[ \hat{T} , \Delta ] = 0 $. Thus, $(\Pi^{k} , V^{\otimes 2k})$ is well defined and forms a representation of $G \times S_{k} \times \mathbb{Z}_{2}$. This is illustrated pictorially in the commutative diagram \ref{Diagram:Modified-Schur-Weyl2}.

\begin{center}\label{Diagram:Modified-Schur-Weyl2}
	\begin{tikzcd}[row sep=2.5em]
		V^{\otimes 2k}  \arrow{rr}{ \hat{T} } \arrow{dr}{ \hat{S}_{\sigma} \otimes \hat{S}_{\sigma} } \arrow{dd}[swap]{ U^{\otimes 2k} } &&
		V^{ \otimes 2k } \arrow[dd,swap,"U^{\otimes 2k}" near start]  \arrow{dr}{ \hat{S}_{\sigma} \otimes \hat{S}_{\sigma} }  \\
		& V^{\otimes 2k} \arrow[rr,crossing over,"\hat{T}" near start,swap] && V^{\otimes 2k}  \arrow{dd}{ U^{\otimes 2k} } \\
		V^{\otimes 2k} \arrow[rr,"\hat{T}" near end] \arrow{dr}[swap]{ \hat{S}_{\sigma} \otimes \hat{S}_{\sigma}  } && V^{ \otimes 2k } \arrow{dr}[swap]{ \hat{S}_{\sigma} \otimes \hat{S}_{\sigma} } \\
		& V^{\otimes 2k} \arrow{rr}[swap]{ \hat{T} } \arrow[uu,<-,crossing over,"U^{\otimes 2k}" near end]&& V^{\otimes 2k}
	\end{tikzcd}
	\begin{figure}[h!]
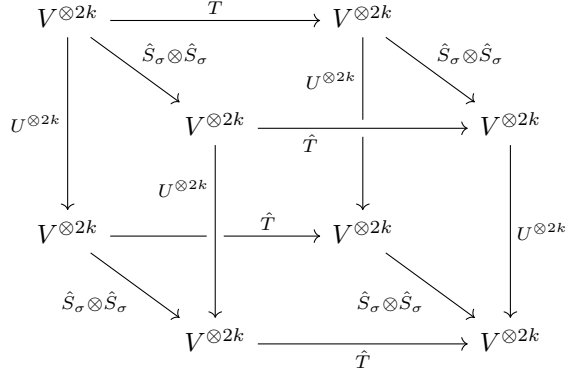

	\caption{ A `cube'-type commutative diagram for the $G \times S_{k} \times \mathbb{Z}_{2}$ representation. The action $(\Pi^{k} , V^{\otimes 2k})$ is well defined and forms a representation of $G \times S_{k} \times \mathbb{Z}_{2}$ only because the $G$ action, the tensor permutation action $S_{k}$ and the $\mathbb{Z}_{2}$ cyclic action are mutually commutative: $[ U^{\otimes 2k} , \hat{S}_{\sigma} \otimes \hat{S}_{\sigma} ] = [ U^{\otimes 2k} , \hat{T} ] = [ \hat{S}_{\sigma} \otimes \hat{S}_{\sigma} , \hat{T} ] = 0$. The observation that $[ U^{\otimes k} , \hat{S}_{\sigma} ] = 0$ is used in the Schur-Weyl duality to show that $k$-fold tensor products of the fundamental representation of $G$ are also representations of $G\times S_{k}$.  }
	\end{figure}
\end{center}


\noindent
Thus, $(\Pi^{k},  V^{\otimes 2k} )$ forms a representation of the group $G \times S_{k} \times \mathbb{Z}_{2} $. The covariance matrix $\Delta$ satisfies the constraint
\begin{align*}
	\forall g \in (G \times S_{k} \times \mathbb{Z}_{2} ), \quad \Pi^{k}(g) \Delta = \Delta \Pi^{k}(g) 
\end{align*}
and so, the matrix $\Delta$ is an element of the endomorphism space of $G \times S_{k} \times \mathbb{Z}_{2}$. We can completely parameterize the endopmorphism space using Schur-Weyl duality. Specifically, the tensor product representation will decompose into irriducible representations of $G \times S_{k} \times \mathbb{Z}_{2}$ as
\begin{align*}
	V^{\otimes 2k}  \cong  \bigoplus_{ \tau \in \hat{G} } \bigoplus_{ \lambda \vdash k } \bigoplus_{ s \in \pm } m^{k}_{ \tau \lambda s }  (\tau, V_{\tau} ) \otimes ( \lambda , V_{\lambda}) \otimes ( s , V_{s} ) 
\end{align*}
where the integers $m^{k }_{\tau \lambda s }$ count the number of irreducible copies of $(\tau, V_{\tau} ) \otimes ( \lambda , V_{\lambda}) \otimes ( s , V_{s} )$ in the representation $( \Pi^{k} , V^{\otimes 2k} )$. The dimension of the endomorphism space is then given by
\begin{align*}
	\dim \text{Hom}_{ G \times S_{k} \times \mathbb{Z}_{2}  }[ ( \Pi^{k} , V^{\otimes k} ) , ( \Pi^{k} , V^{\otimes k} )  ] = \sum_{ \tau \in \hat{G} } \sum_{ \lambda \vdash n } \sum_{ s \in \pm} ( m^{k}_{\tau \lambda s} )^{2}
\end{align*}



\noindent
Furthermore, using the extended Schur lemma \ref{Suppl_Extended_Schur_Lemma}, elements of the endomorpism space are block diagonal
\begin{align*}
	\Phi \in \text{Hom}_{ G \times S_{k} \times \mathbb{Z}_{2}  }[ ( \Pi^{k} , V^{\otimes 2k} ) , ( \Pi^{k} , V^{\otimes 2k} )  ] \implies \Phi = U_{(k,d)}[ \bigoplus_{ \rho \in \hat{G} } \bigoplus_{  \lambda \vdash n } \bigoplus_{  s \in \pm } M^{k}_{ \tau \lambda s  } \otimes \mathbb{I}_{ d_{\tau} }  \otimes \mathbb{I}_{ d'_{\lambda} }    ] U^{\dagger}_{(k,d)}
\end{align*}
where the $M^{k}_{ \tau \lambda s   }$ is a $( m^{k}_{\tau \lambda s} \times m^{k}_{\tau \lambda s} )$-dimensional Hermitian matrix and the unitary matrix $U_{(k,d)}$ is fixed and completely determined. The structure of elements of the endomorphism space  $\text{Hom}_{ G \times S_{k} \times \mathbb{Z}_{2}  }[ ( \Pi^{k} , V^{\otimes k} ) , ( \Pi^{k} , V^{\otimes k} )  ]$ is discussed in the appendix \ref{More_Schur-Weyl_Duality}.

\subsection{Unitary Case}

We consider the case where $G = U(d)$ is the unitary group of dimension $d$. The derivations for the orthogonal and symplectic groups are similar and will be presented in the appendix. A famous result of Weyl in classical group theory \cite{Weyl_1966_Classical} states that the $k$-fold tensor product of the defining representation of the unitary group decomposes as
\begin{align*}
	( \mathbb{C}^{d} )^{ \otimes k }  \cong  \bigoplus_{ \lambda \vdash (k,d) } V_{\lambda} \otimes \lambda 
\end{align*}
where $(k,d)$ denotes all integer partitions of $k$ with less than $d$ summands. Using this decomposition, the representation $( \Pi^{k}, V^{\otimes 2k} )$ is isomorphic to the tensor product representation
\begin{align*}
	( \Pi^{k} , V^{\otimes 2k} )  \cong [ \bigoplus_{ \lambda \vdash (k,d) } V_{\lambda} \otimes \lambda   ] \otimes [ \bigoplus_{ \lambda' \vdash (k,d) } V_{\lambda'} \otimes \lambda'   ]  = \bigoplus_{ \lambda , \lambda' \vdash (k,d) } [ V_{\lambda} \otimes V_{\lambda'} ] \otimes [  \lambda \otimes \lambda' ] 
\end{align*}
The tensor product rules for the unitary group are simple. Let $\lambda \vdash k$ and $\lambda' \vdash k'$. We then have that
\begin{align*}
	V_{\lambda} \otimes V_{\lambda'} = \bigoplus_{ \mu \vdash ( k + k' ) } B^{\mu}_{\lambda \lambda'} V_{\mu}
\end{align*}
where $B^{\mu}_{\lambda \lambda'}$ are the branching rules of the irreducible $\mu \in \hat{S}_{ k + k'} $ under the group restriction $S_{k+k'} \rightarrow S_{k} \times S_{k'}$. Let us also define the tensor product rules of the $S_{k}$ irreducible representations. Let $\lambda \vdash k , \lambda' \vdash k$. Then
\begin{align*}
	\lambda \otimes \lambda' = \bigoplus_{ \mu \vdash k } c_{ \lambda \lambda' }^{ \mu } \mu
\end{align*}
where $c_{ \lambda \lambda' }^{ \mu }$ are called Littlewood-Richardson coefficients. The Littlewood–Richardson coefficients can either be looked up or computed diagrammatically using the Littlewood–Richardson method \cite{james1981representation}.
Using the tensor product rules for unitary irreducible representations and representations of the symmetric group, we have that
\begin{align*}
	( \Pi^{k} , V^{ \otimes 2k } ) \cong \bigoplus_{ \mu \vdash k    } \bigoplus_{ \mu' \vdash k } \bigoplus_{s \in \pm } C^{k}_{ \mu \mu' s } [ V_{ \mu } \otimes \mu' \otimes s ]
\end{align*}
Where we have defined the set of integers,
\begin{align*}
	C^{k }_{ \mu \mu' s } = \sum_{ \lambda  \vdash (k,d) } \sum_{ \lambda' \vdash (k,d) } B^{\mu}_{\lambda \lambda'} c_{ \lambda \lambda' }^{ \mu' } 
\end{align*}
Thus, for each $k$, there exists a fixed unitary matrix $U_{(k,d)}$ such that
\begin{align*}
	( \Pi^{k} , V^{\otimes 2k} ) = U_{(k,d)} [ \bigoplus_{ \mu } \bigoplus_{ \mu' } \bigoplus_{s \in \pm } C^{ks}_{ \mu \mu' }( V_{\mu} \otimes \mu' \otimes s )  ]U_{(k,d)}^{\dagger}
\end{align*}
Again using Schur's lemma, the matrix $\Delta$ can always be written as
\begin{align*}
	\Delta = U_{(k,d)}[ \bigoplus_{ \mu \mu' } \bigoplus_{s \in \pm } M^{ k }_{\mu \mu' s } \otimes \mathbb{I}_{ d_{\mu} d_{\mu'} } ] U_{(k,d)}^{\dagger}
\end{align*}
\noindent
where $M^{k }_{\mu \mu' s}$ is a $C^{k }_{ \mu \mu' s } \times C^{k  }_{ \mu \mu' s } $ matrix. Note that the requirement that $\Delta^{\dagger} = \Delta$ requires that the matrices $(M^{k }_{\mu \mu' s })^{\dagger} = M^{k }_{\mu \mu' s}$ are Hermitian. Lastly, the matrix $\Delta \succ 0 $ is positive definite if and only if each of the matrices $M^{k }_{\mu \mu' s} \succ 0$ are positive definite. The total number of free parameters in the matrix $\Delta$ is given by $\sum_{ \mu \mu' } \sum_{ s \in \pm } ( C^{k  }_{\mu\mu' s } )^{2}$. To summarize we have the following theorem,

\subsection{Main Theorem: Unitary Case } 
\begin{theorem}[ Characterization of $k$-Fold Unitary Endomorpisms ]
	Let $\Phi$ be a $(d^{2k} \times d^{2k})$ positive definite Hermitian matrix. Suppose that
	\begin{align*}
		& \forall U \in U(d), \enspace \forall \sigma \in S_{k},  \enspace \quad [ U^{ \otimes 2k } , \Phi ] =[\hat{S}_{\sigma} \otimes \hat{S}_{\sigma} , \Phi  ] = [ \hat{T} , \Phi ] = 0  
	\end{align*}
	Then the matrix $\Phi$ is completely and uniquely specified by a set of positive definite Hermitian matrices $M_{\mu \mu' s}  = ( M_{ \mu \mu' s } )^{\dagger} \succ 0$. Each $M_{\mu \mu' s}$ matrix is labeled by the integer partitions $\mu, \mu'$ and a sign $s \in \pm$. There exists a fixed unitary matrix $U \in U(d^{2k})$ such that
	\begin{align*}
		\Phi = U [ \bigoplus_{ \mu, \mu' } \bigoplus_{ s \in \pm } M_{ \mu \mu' s } \otimes \mathbb{I}_{ d_{\mu} } \otimes \mathbb{I}_{ d'_{\mu'} } ] U^{\dagger}
	\end{align*}
\end{theorem}
\noindent
where $d_{\mu} = \dim V_{ \mu }$ is the dimension of the $\mu$ irreducible of $U(d)$ and $d'_{\lambda} = \dim \lambda $ is the dimension of the $\lambda$ irreducible of $S_{k}$. Now, the fact that $M_{\mu \mu' s}  = ( M_{ \mu \mu' s } )^{\dagger} \succ 0$ implies that we may write the diagnolization of $M_{ \mu \mu s }$ as
\begin{align*}
	M_{ \mu \mu s } = \sum_{q = 1}^{ C_{ \mu \mu' s} } \Lambda^{q}_{ \mu \mu' s } | \phi^{q}_{ \mu \mu' s } \rangle \langle \phi^{q}_{ \mu \mu' s } |
\end{align*}
where the eigenvalues $\Lambda^{q}_{ \mu \mu' s }  > 0$ and the $C_{\mu \mu' s}$-dimensional vectors $| \phi^{q}_{ \mu \mu' s } \rangle$ are orthonormal with
\begin{align*}
	\forall q,q' \in \{ 1, 2, ..., C_{\mu \mu' s} \} , \quad \langle \phi^{q'}_{ \mu \mu' s } | \phi^{q}_{ \mu \mu' s } \rangle  = \delta^{qq'}
\end{align*}

\section{ Bipartite Unitary Ensemble }

\noindent
We consider the $k=2$ case. The bipartite case is of great importance. To begin, let us calculate the all important coefficients $C^{(2)}_{\mu\mu'}$. Assuming that $d>1$, we have that
\begin{align*}
	\mu \vdash 4, \enspace \mu' \vdash 2, \quad C^{(2)}_{ \mu \mu' } = \sum_{ \lambda \vdash 2 } \sum_{ \lambda' \vdash 2 } B_{\lambda\lambda}^{\mu} c^{\lambda \lambda'}_{\mu'}
\end{align*}
The non-zero Littlewood-Richard coefficients for $S_{2} \cong \mathbb{Z}_{2}$ are easy to compute,
\begin{align*}
	c^{ ++ }_{+} = 1, \quad c^{+-}_{-} = 1, \quad c^{--}_{+} = 1
\end{align*}
and the non-zero branching rules for $S_{4} \rightarrow S_{2} \times S_{2}$ are computed in \ref{Rules for bipartite},
\begin{align*}
	& B^{++}_{(4)} = 1 \quad B^{--}_{(1,1,1,1)} = 1, \quad B^{++}_{(2,2)} = B^{--}_{(2,2)} = 1 \\
	& B^{++}_{(2,1,1)} = B^{+-}_{(2,1,1)}=B^{-+}_{(2,1,1)} = 1, \quad B^{--}_{(3,1)} = B^{+-}_{(3,1)} = B^{-+}_{(3,1)} = 1
\end{align*}
Thus, the dimensions of the matrices appearing in the covariance matrix are given by
\begin{align*}
	& C^{(2)}_{(4)+} = 1, \quad C^{(2)}_{(1,1,1,1),+} = 1, \quad C^{(2)}_{(2,2),+} =  C^{(2)}_{(2,1,1),+} = 1 \\
	& C^{(2)}_{(2,1,1),-} = 2, C^{(2)}_{(3,1),+} = 1, \quad C^{(2)}_{(3,1),-} = 2 \\
\end{align*}
\noindent
The endomorpism space is then isomorphic to,
\begin{align*}
	\Hom_{ U(d) \times S_{2} }[ ( \Pi^{2} , ( \mathbb{C}^{d} )^{\otimes 4} )  ] \cong \lambda_{1} \oplus \lambda_{2} \oplus \Lambda_{1} \oplus \lambda_{3} \oplus \Lambda_{2} \oplus \Lambda_{4} \oplus \Lambda_{3}
\end{align*}
where $\lambda_{i}$ are numbers and $\Lambda_{i}$ are $2\times 2$ matrices. The dimension of the endomorpism space is then given by,
\begin{align*}
	\dim \Hom_{ U(d) \times S_{2} \times \mathbb{Z}_{2} }[ ( \Pi^{2} , ( \mathbb{C}^{d} )^{\otimes 4} )  ] = 4 \cdot 1^{2} + 3 \cdot 2^{2} = 16
\end{align*}
and the endomorpism space has $16$ free parameters, independent of the dimension $d$. When we enforce positively of the convarince matrix $\Delta \succ 0$, we have that
\begin{align*}
	& \lambda_{1}, \lambda_{2}, \lambda_{3}, \lambda_{4} \geq 0 \\
	& \Lambda_{1} = \Lambda^{T}_{1} \succ 0, \Lambda_{2} = \Lambda^{T}_{2} \succ 0, \Lambda_{3} = \Lambda^{T}_{3} \succ 0
\end{align*}
Thus the total number of real free parameters is $ 4 + 3\cdot 3=13$. Thus, the most general unitary random matrix ensemble that is two-fold unitary $U(d)^{\otimes 2}$ invariant has $13$ free parameters, independent of the dimension $d$.

\section{Physical Models}\label{Appendix:Physical_Models}

In order to motivate our proposed distributions, we consider a set of physical models that exhibit invariance of measure under $k$-fold normal transformation but not invariance under conjugation by generic unitary matrix. The $k$-fold invariance is intimately related to the existence of gapless modes.

	\subsection{$k$-Fold Spin Models}
	\noindent
	
	We show that the property of $k$-fold invariance arises quite naturally in disordered Heisenberg models. In the noiseless limit, the Heisenberg model of interacting spins has an exact continuous symmetry and is thus $k$-fold invariant. When noise is added, this symmetry is broken. However, if the noise is isotropic, the family of random matrices will still be $k$-fold invariant. These examples also illustrate the fact that $k$-fold invariance is related to breaking of symmetry with isotropic noise.

	\subsubsection{$k$-Fold Unitary Invariance }
	\noindent
	Consider the standard $SU(2)$ spin operators $S = (S^{x},S^{y} ,S^{z} )$ satisfying the commutation relation
	\begin{align}\label{spin_algebra}
		[ \hat{S}^{\alpha} , \hat{S}^{\beta}  ] = i \epsilon^{\alpha\beta \gamma} \hat{S}^{\gamma}
	\end{align}
	and transforming in the $\frac{1}{2}$-irreducible representation of $SU(2)$. We are always free to redefine the coordinate system for each spin.
	Specifically, the change of basis $S^{i} \rightarrow U S^{i} U^{\dagger}$, where $U$ is any unitary matrix, preserves the commutation relations \ref{spin_algebra}. Now, let $\mathcal{G} = ( V , E )$ be a graph. Let $\hat{n}_{i} \in \mathbb{R}^{3}$ be a set of random vectors that are drawn with uniform angular distribution in $\mathbb{R}^{3}$. As an example, one could consider a Gaussian density for the $\hat{n}_{i}$,
	\begin{align*}
		\text{Pr}[ \hat{n} ] d\hat{n} \propto \exp( \frac{-1}{2} || \hat{n} ||^{2}_{2} ) d\hat{n}
	\end{align*}
	Because the magnetic field is drawn isotropically in $\mathbb{R}^{3}$, the vector valued random variable $\hat{n}_{i}$ satisfies,
	\begin{align*}
		\forall U \in SU(2), \quad \hat{n}_{i} \cdot U \hat{S}_{i} U^{\dagger} = \hat{n}_{i} \cdot \hat{S}_{i} 
	\end{align*}
	Now, consider the disordered Heisenberg model defined on the graph $G$,
	\begin{align}\label{Disordered_Heisenberg}
		H = \sum_{ ij \in E } J_{ij} \hat{S}_{i} \cdot \hat{S}_{j} + \sum_{ i \in V } \hat{n}_{i} \cdot \hat{S}_{i}
	\end{align}
	where $J_{ij}$ are some fixed constants which measures the coupling along the $ij$-th edge. The Hamiltonians in \ref{Disordered_Heisenberg} are matrix valued random variables drawn from a probability distribution since each $\hat{n}_{i}$ is a random variable. 
	
	With zero external magnetic fields $\hat{n}_{i} = 0$, the Heisenberg model has a local $SU(2)$ degree of freedom due to the fact that a rotation of each spin operator by the same amount does not change the energy. Because the noise $\hat{n}_{i}$ is drawn isotropically, this transformation holds at the level of random variables. Now, consider the transformation that redefines the local Hilbert space by the same unitary matrix
	\begin{align*}
		\hat{S}_{i} \rightarrow U \hat{S}_{i} U^{\dagger}
	\end{align*}
	under this transformation,
	\begin{align*}
		H \rightarrow H' = \sum_{ ij \in E } J_{ij} U (\hat{S}_{i} \cdot \hat{S}_{j} ) U^{\dagger}
	\end{align*}
	Using the relation $UU^{\dagger}= \mathbb{1} = U^{\dagger}U$ and the fact that unitary matrices on different tensor product subspaces commute, we can rewrite this expression as,
	\begin{align*}
		H' =  \sum_{ ij \in E } J_{ij} U (\hat{S}_{i} \cdot \hat{S}_{j} ) U^{\dagger} = \sum_{ ij \in E } J_{ij} (U^{\otimes k} ) (\hat{S}_{i} \cdot \hat{S}_{j} ) ( U^{\otimes k} )^{\dagger} =  (U^{\otimes k} ) [ \sum_{ ij \in E } J_{ij} (\hat{S}_{i} \cdot \hat{S}_{j} ) ] ( U^{\otimes k} )^{\dagger} =  (U^{\otimes k} ) H ( U^{\otimes k} )^{\dagger}
	\end{align*}
	Thus, under the $k$-fold normal transformation, the probability density satisfies $P(H)dH = P(H')dH'$. It should be noted that this probability distribution is not invariant under the more general transformation $H \rightarrow VHV^{\dagger}$ where $V \in U(d^{k} )$ is a arbitrary unitary matrix.

	\subsubsection{ $k$-Fold Gaussian Orthogonal Models }
	
	Quantum mechanical systems that are time reversal invariant can be represented with symmetric Hamiltonian \cite{Landau_1991_Quantum}. We give an example of a natural disordered Hamiltonian that is invariant under $k$-fold Orthogonal transformation. Consider the $SO(3)$ spin operators $S = (S^{x},S^{y} ,S^{z} )$ satisfying the commutation relation
	\begin{align}\label{spin_algebra_II}
		[ \hat{S}^{\alpha} , \hat{S}^{\beta}  ] = i \epsilon^{\alpha\beta \gamma} \hat{S}^{\gamma}
	\end{align}
	Consider the $\ell=1$ representation ( i.e. the $\ell=1$ spin sector). Then, $S = (S^{x},S^{y} ,S^{z} )$ are real $ 3\times3 $ matrices. Let $O_{ij}= O^{T}_{ji} \in O(n)$ be a set of orthogonal matrices, for example drawn randomly from a Langevin distribution \cite{mardia2009directional}. This is a natural model as the Langevin distribution is the maximum entropy distribution with fixed first moment on $O(d)$ \cite{mardia2009directional,Jaynes_1982}. Suppose that the distribution of $O_{ij}$ is isotropic. Then,
	\begin{align*}
		\forall R \in O(3), \quad RO_{ij}R^{T} = O_{ij}
	\end{align*}
	must hold at the level of random variables as an isotropic distribution has no preferential basis. Consider a disordered $O(3)$ model of the form
	\begin{align*}
		\hat{H} = \sum_{ ij } J_{ij} \vec{S}_{i}^{T} O_{ij} \vec{S}_{j} \end{align*}
	where the $J_{ij} \in \mathbb{R}$ measure the alignment affinity between $i$-th and $j$-th spin and $J_{ij}$ are drawn randomly from some distribution. In this model it is energetically favorable for the $i$-th and $j$-th spin vectors to form a relative angle of $O_{ij}$. In the noiseless case $O_{ij} = \mathbb{1}_{3}$, the disordered $O(3)$ model has a $O(3)$ symmetry. Specifically, for any orthogonal matrix $O \in O(3)$ the transformation $ \vec{S}_{k} \rightarrow O \vec{S}_{k} O^{T}$ preserves both the commutation relations \ref{spin_algebra_II} are preserved and the Hamiltonian. Note that the addition of the random noise terms $O_{ij}$ break the $O(3)$-symmetry, but because of the isotropy properties of Langevin distributions this relation still holds in the noisy case. We can write this model as
	\begin{align*}
		\hat{H}  = \sum_{ij} J_{ij} \vec{S}^{T}_{i} O_{ij} \vec{S}_{j} 
	\end{align*}
	Under the $ \vec{S} \rightarrow O \vec{S} O^{T} $ symmetry transformation, we have that $\hat{H}$ is invariant under $k$-fold normal transformation by an $O(n)$ degree of freedom.

	
	\subsection{ Quantum Double Fold Models }
	
	Another situation where $k$-fold invariance appears naturally is in gauge theories \cite{Lahtinen_2017}. Quantum double fold models \cite{pachos2012introduction} are a group theoretic generalization of the Kitaev toric code \cite{Kitaev_2003Fault}. Specifically, for any finite group $G$, the quantum double fold models describe interactions between local Hilbert spaces of size $|G|$. We show that quantum double fold models satisfy the $k$-fold $G$-invariance property. This is intuitively obvious, as a change of gauge leaves the Hamiltonian invariant.
	
	\subsubsection{Quantum Double Fold Models}
	We review the quantum double fold models \cite{pachos2012introduction}. Let $G$ be a finite group. The quantum double fold models are defined as the following: Let $\mathcal{H}$ be a vector space of orthogonal states
	\begin{align*}
		\mathcal{H} = \{ \enspace | g \rangle \enspace | \enspace \forall g \in G \enspace  \}
	\end{align*}
	with orthogonal inner product $\langle g| g' \rangle = \delta_{gg'}$. Define the operators
	\begin{align*}
		& T^{+}_{g} |z\rangle = |gz \rangle, \quad T^{-}_{g} |z\rangle = |z g^{-1} \rangle \\
		& P^{+}_{h} |z \rangle = \delta_{h,z}|h\rangle, \quad P^{-}_{h} |z \rangle = \delta_{h^{-1},z}| h^{-1} \rangle 
	\end{align*}
	so that $(T^{+} , \mathcal{H} )$ and $(T^{-} , \mathcal{H} )$ form a unitary representation of the group $G$. Consider the the Hilbert space $\otimes_{i=1}^{k} \mathcal{H}$ which consists of $k$ copies of $\mathcal{H}$. The vertex and plaquette operators are then defined on $\otimes_{i=1}^{k} \mathcal{H}$. The vertex and plaquette operators are given by
	\begin{align*}
		& \text{Vertex: } A(v) = \sum_{g\in G} A_{g}(v) = \sum_{g\in G} T^{+}_{g}T^{+}_{g}T^{-}_{g}T^{-}_{g} \\
		& \text{Plaquette: }B(p) =  \sum_{ h_{1}h_{2}h_{3}h_{4} = 1 } P^{+}_{h_{1}}P^{+}_{h_{2}}P^{-}_{h_{3}}P^{-}_{h_{4}}
	\end{align*}
	The Hamiltonian is then defined as
	\begin{align*}
		H = -\sum_{v} A(v) - \sum_{p} B(p)
	\end{align*}
	Now, note that the choice of labeling of basis in $\mathcal{H}$ is arbitrary. Specifically, we are always free to redefine $| g \rangle \rightarrow | \phi(g) \rangle$ where $\phi$ is an automorphism of the group $G$. Let $q \in G$ be a group element of $G$. Let us consider the relabeling corresponding to left multiplication by a fixed element $q \in G$,
	\begin{align*}
		| g \rangle \rightarrow  T^{+}_{q} | g \rangle = | q g \rangle 
	\end{align*}
	Again, this is just a relabeling of the state space $\mathcal{H}$, and should not change any physical quantities. Under this transformation, we have that
	\begin{align*}
		& T^{+}_{g} \rightarrow T^{+}_{q} T^{+}_{g} T^{+}_{q^{-1}} = T^{+}_{qgq^{-1}}, \quad T^{-}_{g} \rightarrow T^{+}_{q} T^{-}_{g} T^{+}_{q^{-1}} = T^{-}_{g} \\
		& P^{+}_{g} \rightarrow T^{+}_{q} P^{+}_{g} T^{+}_{q^{-1}}, \quad  P^{-}_{g} \rightarrow T^{+}_{q} P^{-}_{g} T^{+}_{q^{-1}}
	\end{align*}
	Now, how do the vertex operators $A(v)$ and plaquette $B(p)$ operators transform? We have that
	\begin{align*}
		\text{Vertex Transformation: } A(v) \rightarrow \sum_{g \in G } T^{+}_{ qgq^{-1} } T^{+}_{ qgq^{-1} } T^{-}_{ qgq^{-1} } T^{-}_{ qgq^{-1} }  = \sum_{qgq^{-1} \in G } T^{+}_{ qgq^{-1} } T^{+}_{ qgq^{-1} } T^{-}_{ qgq^{-1} } T^{-}_{ qgq^{-1} }  = A(v)
	\end{align*}
	relabeling this summation $g \rightarrow qgq^{-1}$, we see that $A(v) \rightarrow A(v)$ and the vertex operator is invariant. The plaquette operator $B(p)$ is similarly invariant. Under transformation $|g\rangle \rightarrow |qg\rangle$,
	\begin{align*}
		\text{Plaquette Transformation: } B(p) \rightarrow \sum_{ h_{1}h_{2}h_{3}h_{4} = 1 } (P^{+}_{ qh_{1}q^{-1} })(P^{+}_{ qh_{2}q^{-1} })(P^{+}_{ qh_{3}q^{-1} })(P^{+}_{ qh_{4}q^{-1} }) = \sum_{ h_{1}h_{2}h_{3}h_{4} = 1 } P^{+}_{ h_{1} }P^{+}_{ h_{2} }P^{+}_{ h_{3} }P^{+}_{ h_{4} } = B(p)
	\end{align*}
	relabeling the summation indices $h_{i} \rightarrow qh_{i}q^{-1}$, the product $h_{1}h_{2}h_{3}h_{4} = 1$ is unchanged as
	\begin{align*}
		h_{1}h_{2}h_{3}h_{4} \rightarrow (qh_{1}q^{-1})(qh_{2}q^{-1})(qh_{3}q^{-1})(qh_{4}q^{-1}) = q \underbrace{ h_{1}h_{2}h_{3}h_{4} }_{ = 1} q^{-1} = qq^{-1}= 1
	\end{align*}
	we see that the plaquette operator $B(p)$ is also invariant. Thus, under the relabeling $|g\rangle \rightarrow |qg\rangle$, both the vertex operator and the plaquette operator are invariant
	\begin{align*}
		A(v) \rightarrow A(v), \quad B(p) \rightarrow B(p)
	\end{align*}
	and the double fold model Hamiltonian is invariant under global labeling. Thus, in the original basis the Hamiltonian is invariant under $k$-fold $G$-transformation on the Hilbert space $\otimes_{i=1}^{k} \mathcal{H}$. Intuitively, this is obvious, as this relabeling $|g\rangle \rightarrow |hg\rangle$ does not change the physics of the double fold model in any way.
	
	\subsubsection{Unitary Double Fold Model}
	
	The quantum double fold models with group $G$ are $k$-fold $G$-invariant. We can generalize these models to the group $U(d)$. Analogous to the quantum double fold model, let us define the Hilbert spaces $\mathcal{H}$ to be vectorizations of $U(d)$ matrices transforming in the fundamental (i.e. $d$-dimensional) representation,
	\begin{align*}
		\mathcal{H} = \{ \enspace | V \rangle \enspace | \enspace | V \rangle = \text{Vec}(V) , \enspace V \in U(d) \enspace  \}
	\end{align*}
	The inner product between two states $| V \rangle \in \mathcal{H}$ and $| V' \rangle \in \mathcal{H} $ is defined as $ \langle V | V' \rangle = \frac{1}{d}\text{Tr}[ V^{\dagger}V' ] $. We can then define the operators
	\begin{align*}
		\hat{T}^{+}(U) | V \rangle = |UV \rangle, \quad \hat{T}^{-}(U) | V \rangle = | VU^{-1} \rangle
	\end{align*}
	so that $\hat{T}^{\pm}: U(d) \rightarrow \Hom[ \mathcal{H}, \mathcal{H} ]$. These operators are both norm preserving and invertable and therefore unitary. Let us define the operators acting on $\mathcal{H} \otimes \mathcal{H}$ as
	\begin{align*}
		\hat{T}_{ij} = \int_{U(d)} dU \text{ } \hat{T}_{i}^{+}(U) \otimes \hat{T}_{j}^{-}(U) = \mathbb{E}_{U}[ \hat{T}_{i}^{+}(U) \otimes \hat{T}_{j}^{-}(U)  ]
	\end{align*}
	where $\mathbb{E}_{U}[ \enspace \cdot \enspace ]$ denotes the expectation with respect to the Haar measure \cite{Weyl_1966_Classical}. There is a closed form expression for $\hat{T}_{ij}$ derived in (ref appendix). 
	The matrix elements of the operator $\hat{T}_{ij}$ are given by,
	\begin{align*}
		\langle V_{1} , V_{2} |  \hat{T}_{ij} | V_{1}' , V_{2}' \rangle =  \alpha \langle V_{1} | V'_{1} \rangle \langle V_{2} | V'_{2} \rangle + \beta d \langle V_{1} | V'_{2} \rangle \langle V'_{1} | V_{2} \rangle
	\end{align*}
	where $\alpha$ and $\beta$ are some constants. We can then define the Hamiltonian 
	\begin{align*}
		\mathcal{H} = \sum_{ ij } \hat{T}_{ij}
	\end{align*}
	Note that under the gauge transformation $|V \rangle \rightarrow \hat{T}^{+}(U) | V \rangle = |UV \rangle$ the $\hat{T}_{ij}$ terms are invariant. Specifically, we have that
	\begin{align*}
		\forall V \in U(d), \quad  [ \hat{T}^{+}_{i}(V) \otimes \hat{T}^{+}_{j}(V) ]  \hat{T}_{ij} =  \int_{U(d)}dU\text{ }[ \hat{T}^{+}_{i}(V) \hat{T}^{+}_{i}(U) \otimes \hat{T}^{+}_{j}(V) \hat{T}^{-}_{j}(U) ]= \int_{U(d)}dU\text{ }[ \hat{T}^{+}_{i}(VU) \otimes \hat{T}^{+}_{j}(V) \hat{T}^{-}_{j}(U) ]
	\end{align*}
	Using the commutativity $ \hat{T}^{+}(U)\hat{T}^{-}(V) = \hat{T}^{-}(U)\hat{T}^{+}(V)$, we have that
	\begin{align*}
		V\in U(d), \quad [\hat{T}^{+}_{i}(V)\otimes \hat{T}^{-}_{j}(V)] \hat{T}_{ij} = \hat{T}_{ij}[\hat{T}^{+}_{i}(V)\otimes \hat{T}^{-}_{j}(V)]
	\end{align*}
	Thus, the unitary double fold model has unitary $k$-fold invariance. Note that we can also define higher order terms,
	\begin{align*}
		\hat{T}_{ijkl} = \int_{ U(d) }dU \text{ } \hat{T}^{+}_{i}(U) \otimes \hat{T}^{+}_{j}(U) \otimes \hat{T}^{-}_{k}(U) \otimes \hat{T}^{-}_{l}(U)
	\end{align*}
	which will again be $k$-fold invariant.

	\subsection{$k$-Fold Bosonic Models}
	
	\noindent
	Suppose that we have a physical system described by set of $d$ single particle orbitals 
	\begin{align*}
		\mathcal{S} = \text{Span}\{ |\phi_{s} \rangle \}_{s=1}^{d} 
	\end{align*}
	The choice of basis functions $|\phi_{s} \rangle$ is a choice of coordinate system on $\mathcal{S}$. Specifically, we are always free to transform the basis functions $|\phi_{s} \rangle \rightarrow \sum_{s'=1}^{d} U_{ss'}|\phi_{s'} \rangle $ where $U$ is some unitary matrix. We wish to describe a set of theories that depend only on the intrinsic geometry of the space $\mathcal{S}$. Let $G$ be a compact group, either Lie or finite. We will assume that $G$ has action on the space $\mathcal{S}$ so that $ ( \rho , \mathcal{S} ) $ is a $d$-dimensional $G$ representation. Furthermore, suppose that we have $k$ identical copies of $\mathcal{S}$, so that the Hilbert space of our system is then described by
	\begin{align*}
		\mathcal{H} = \underbrace{ \mathcal{S} \otimes \mathcal{S} \otimes ... \otimes \mathcal{S} }_{k-times} = \mathcal{S}^{ \otimes k }
	\end{align*}
	We wish to describe theories that depend only on the relative geometries of each of the copies of $\mathcal{S}$. This means that under an identical change of basis in each of the $\mathcal{S}$, the physics of our theory should not change. Using the second quantization formalism \cite{Landau_1991_Quantum}, our system can be described by bosonic creation and annihilation operators carrying a $d$-dimensional internal index $s$ and a $k$-dimensional external index $i$,
	\begin{align*}
		[ \hat{b}_{i,s} , \hat{b}_{j,s'} ] = 0, \quad [ \hat{b}_{i,s} , \hat{b}^{\dagger}_{j,s'} ] = \delta_{ij} \delta_{ss'} , \quad [ \hat{b}^{\dagger}_{i,s} , \hat{b}^{\dagger}_{j,s'} ] = 0 
	\end{align*}
	where the operator $\hat{b}^{\dagger}_{i,s}(\hat{b}^{\dagger}_{i,s}) $ creates(destroys) a particle in the state $| \phi_{s} \rangle $ on the $i$-th copy of the space $\mathcal{S}$. The Hilbert space of this system is spanned by the states
	\begin{align*}
		| m^{ (1) },m^{ (2) },...,m^{ (k) } \rangle \propto \prod_{i=1}^{k} (\hat{b}_{i1}^{\dagger})^{ m_{i1} }(\hat{b}_{i2}^{\dagger})^{ m_{i2} } ... (\hat{b}_{id}^{\dagger})^{ m_{id} } | \text{vac} \rangle = \prod_{i=1}^{k} \prod_{j=1}^{d} (\hat{b}_{ij}^{\dagger})^{ m_{ij} } | \text{vac} \rangle
	\end{align*}
	where each of the $m_{i}=m_{i1}m_{i2}...m_{id}$ are tuples of $d$ positive integers. Under a $G$ transformation on $\mathcal{S}$, the bosonic operators transform as a representation of $G$,
	\begin{align*}
		\forall g \in G, \quad g \cdot \hat{b}_{i,s} = \sum_{s'} \rho(g)_{s,s'} \hat{b}_{i,s'}, \quad g \cdot \hat{b}^{\dagger}_{i,s} = \sum_{s'} \rho^{\dagger}(g)_{s,s'} \hat{b}^{\dagger}_{i,s'}
	\end{align*}
	where $( \rho , \mathbb{C}^{d} )$ is some $d$-dimensional representation of the group $G$. Let us define the site number operators
	\begin{align*}
		\hat{n}_{i} =\sum_{s} \hat{n}_{i,s} =  \sum_{s} \hat{b}^{\dagger}_{i,s}\hat{b}_{i,s} 
	\end{align*}
	so that the operator $\hat{n}_{i}$ counts the total number of particles on site $i$. The site number operator is a physical quantity and is invariant under $G$ transformation with $g \cdot \hat{n}_{i} = \hat{n}_{i}$. We would like to understand how states transform under a relabeling of the bosonic operators. Let us consider symmetry action on states $g \cdot |m_{1}m_{2}...m_{k} \rangle$. Consider the transformation at the $i$-th site,
	\begin{align*}
		\prod_{j=1}^{d} (\hat{b}_{ij}^{\dagger})^{ m_{ij} } \rightarrow \prod_{j=1}^{d} ( \sum_{j'} \rho(g)_{jj'} \hat{b}_{ij'}^{\dagger} )^{m^{(i)}_{j} } = \sum U(g)_{ m_{i1}m_{i2}...m_{id}, m'_{i1}m'_{i2}...m'_{id} } \prod_{j=1}^{d} (\hat{b}_{ij}^{\dagger})^{ m'_{ij} }
	\end{align*}
	Where $U(g)$ has $G$ action on states given by,
	\begin{align*}
		|m_{i1}m_{i2}...m_{id} \rangle \rightarrow \sum_{m'_{ij}} U(g)_{ m_{i1}m_{i2}...m_{id} ,m'_{i1}m'_{i2}...m'_{id} } |m'_{i1}m'_{i2}...m'_{id} \rangle
	\end{align*}
	so that $(U, \mathcal{S}^{\otimes k})$ is a $G$-representation. Thus, under a $G$ transformation, we have that states transform as
	\begin{align*}
		\forall g \in G, \enspace \forall | \Psi \rangle \in \mathcal{H} \quad g \cdot | \Psi \rangle \rightarrow [ U(g) \otimes U(g) \otimes ... \otimes U(g) ] | \Psi \rangle  = U(g)^{\otimes k} | \Psi \rangle
	\end{align*}
	Now, consider Hamiltonians with of the form
	\begin{align}\label{QuadraticBosonHamiltonian}
		\hat{H} =  \sum_{ ij \in E }  T^{ij}_{ss'} \hat{b}^{\dagger}_{i,s}\hat{b}_{j,s'}   
	\end{align}
	where $T^{ij}_{ss'}$ is a complex valued scalar random variable characterizing transition amplitudes. Note that this terms of this form preserves site particle number as $[ \hat{n}_{i} , \hat{H} ] = 0  $. Now, under a uniform gauge transformation, we have that
	\begin{align*}
		g \cdot \hat{H} \rightarrow \sum_{ij \in E} \hat{b}^{\dagger}_{i,t} \rho^{\dagger}_{ts}(g) T^{ij}_{ss'} \rho_{s't'}(g) \hat{b}_{j,t'} 
	\end{align*}
	Thus, if we require that the $d\times d$ random Hermitian matrix $T^{ij} = ( T^{ij} )^{\dagger}$ satisfies
	\begin{align*}
		\forall g \in G, \quad \rho(g) T^{ij} \rho(g)^{\dagger} = T^{ij} \implies T^{ij} \in \Hom_{G}[ \rho , \rho ]
	\end{align*}
	Then the Hamiltonian random variable $\hat{H}$ is invariant under the $G$ transformation. We can similarly define interaction terms
	\begin{align*}
		H_{int} =  \sum_{ijkl} \sum_{ss'tt'} C^{ijkl}_{ss'tt'}\hat{b}^{\dagger}_{is}\hat{b}^{\dagger}_{js'}\hat{b}_{kt}\hat{b}_{lt'}
	\end{align*}
	Then, under a $G$-transformation, we have that,
	\begin{align*}
		\forall g \in G, \quad g \cdot H_{int} =  \sum_{ijkl} \sum_{ss'tt'} C^{ijkl}_{ss',tt'}\rho_{su}^{\dagger}(g)\rho_{s'u'}^{\dagger}(g)\rho_{tv}(g)\rho_{t'v'}(g) \hat{b}^{\dagger}_{iu}\hat{b}^{\dagger}_{ju'}\hat{b}_{kv}\hat{b}_{lv'}
	\end{align*}
	This term is $G$-invariant if and only if
	\begin{align*}
		\forall g \in G, \quad [ \rho(g) \otimes \rho(g) ] C^{ijkl}  = C^{ijkl} [ \rho(g) \otimes \rho(g) ] \implies C^{ijkl} \in \Hom_{G}[ \rho \otimes \rho , \rho \otimes \rho ]
	\end{align*}
	so that $C^{ijkl}$ is an endomorpism of the $ \rho \otimes \rho $ representation. To summarize: let $\mathcal{S} = \text{Span}[ \{ | \Phi_{i} \rangle  \}_{i=1}^{n} ]$ describe a set of single particle orbitals. Second quantized systems on $\mathcal{H} = \bigotimes_{i=1}^{k} \mathcal{S} $ which depend only on the relative geometry of each of the $\mathcal{S}$ spaces have Hamiltonian operators given by
	\begin{align*}
		& \text{Non-Interacting: } \hat{H} = \sum \hat{b}^{\dagger}_{is} T^{ij}_{ss'}\hat{b}_{js'} \text{ with } T^{ij} \in \Hom_{G}[ \rho , \rho ] \\ 
		& \text{Interacting: } \hat{H} = \sum C^{ijkl}_{ss'tt'}  \hat{b}^{\dagger}_{is} \hat{b}^{\dagger}_{js'} \hat{b}_{kt} \hat{b}_{lt'} \text{ with }  C^{ijkl} \in \Hom_{G}[ \rho \otimes \rho , \rho \otimes \rho ] 
	\end{align*}
	Terms that do not preserve particle number satisfy similar constraints.

	\subsubsection{Specific Example: Harmonic Oscillator}
	
	As a more specific example, let us consider a system of $N$-interacting three-dimensional harmonic oscillators. The operators $\hat{b}_{i,0,0}$ are of degeneracy 1 and transforms in the trivial representation of $SO(3)$. Similarly, the triple $\hat{b}_{i,1,-}, \hat{b}_{i,1,0}, \hat{b}_{i,1,1}$ form the $\ell = 1$ representation of $SO(3)$. In general, operators can be written as $\hat{b}_{i,\ell k}$ transforming in the $\ell$-representation of $SO(3)$,
	\begin{align*}
		g \cdot \hat{b}_{i,\ell k} = \sum_{k=-\ell}^{\ell} D^{\ell}(g)_{kk'} \hat{b}_{i,\ell k'}
	\end{align*}
	where $D^{\ell}$ are the Wigner $D$-matrices. The allowed $SO(3)$-invariant quadratic terms are then given by
	\begin{align*}
		\hat{H} =  \sum_{ij=1}^{N} \sum_{\ell=0}^{\infty} z_{\ell, ij  }  \sum_{k=-\ell}^{\ell} \hat{b}^{\dagger}_{i, \ell k } \hat{b}_{j, \ell k}
	\end{align*}
	where $z_{\ell, ij} = \bar{z}_{\ell, ji} \in \mathbb{C}$.

	\section{ Field Theory Description}\label{Field_Theory_Description}
	The result in \ref{Charecterization_Lemma} gives a complete description of a set of natural matrix ensembles. By property II, the derived probability distributions are Gaussian and all matrix element correlation functions can be computed using Isserlis' theorem. The standard Gaussian Ensembles are equivalently described as non-interacting $(0+0)$-dimensional field theories. From a field theory perspective, Gaussian matrix probability distributions are non-interacting theories, as all correlation functions can be computed exactly. Specifically, in the $k$-fold matrix ensemble, any $n$-point correlation of the form,
	\small
	\begin{align*}
		\langle H_{ \underbrace{i_{11}i_{12}...i_{1k}}_{i_{1}}  \underbrace{j_{11}j_{12}...j_{1k}}_{j_{1}} } H_{ \underbrace{i_{21}i_{22}...i_{2k}}_{i_{2}}  \underbrace{j_{21}j_{22}...j_{2k}}_{j_{2}} } ... H_{ \underbrace{i_{n1}i_{n2}...i_{nk}}_{i_{n}}  \underbrace{j_{n1}j_{n2}...j_{nk}}_{j_{n}} }  \rangle \propto \int \prod_{i \leq j} dH_{ij} \text{ } H_{i_{1}j_{1}}H_{i_{2}j_{2}}...H_{i_{n}j_{n}} \exp( \frac{-1}{2} \Vec{H}^{\dagger} \Delta \Vec{H} )
	\end{align*}
	\normalsize
	can be computed exactly as all integrals are Gaussian. The diagrammatic rules are slightly more complicated than a standard Gaussian field theory. This can be represented diagrammatically in Feynman diagrams. Unlike standard Feynman diagrams, each `particle' index carries $k$ tensor indices.

	\subsubsection{ Eigenvalue-Entanglement Interactions }\label{Eigenvalue-Entanglement Interactions}
	\noindent
	One novel feature of the proposed $k$-fold ensembles is the existence of terms that couple entanglement spectra to eigenvalue spectra. Let us illustrate this by considering the $k=2$ bipartite case. To begin, consider a standard quartic interaction term like $\text{Tr}[ \hat{H}^{4} ]$ which depends only on the eigenvalue spectrum of the matrix $\hat{H}$. Consider the diagonilization of $\hat{H}$,
	\begin{align}\label{Diag_form_H_II}
		\hat{H} = \Psi \Lambda \Psi^{\dagger} = \sum_{ m } \Lambda_{m} \Psi_{m}\Psi^{\dagger}_{m}
	\end{align}
	then, $\text{Tr}[ \hat{H}^{4} ] = \text{Tr}[ \Lambda^{4} ] = \sum_{m} \Lambda^{4}_{m} $ which is completely independent of the form of the eigenvectors $\Psi$. Now, let $\hat{S}$ be the swap operator on the vector space with $\hat{S}|ij \rangle = |ji \rangle $. Now, let
	\begin{align*}
		\hat{H}_{S} = \hat{S}\hat{H}\hat{S}^{\dagger}
	\end{align*}
	\noindent
	Consider the quadratic interaction term
	\begin{align*}
		\text{Tr}[  (\hat{H}_{S}\hat{H})^{\dagger} (\hat{H}_{S}\hat{H})  ] = \text{Tr}[ \hat{H}^{\dagger}_{S} \hat{H}_{S} \hat{H}^{\dagger} \hat{H}  ] = \text{Tr}[ \hat{H}_{S}^{2} \hat{H}^{2} ]
	\end{align*}
	this term is forbidden under $U(d^{2})$ unitary invariance, but allowed under $2$-fold $U(d)$ unitary invariance. We can write down the $m$-th Schmidt decomposition of the eigenvectors of $\hat{H}$,
	\begin{align*}
		\Psi_{m} = \sum_{n} \lambda_{mn} | \psi^{L}_{mn} ,  \psi^{R}_{mn} \rangle
	\end{align*}
	where $| \psi^{L}_{mn} ,  \psi^{R}_{mn} \rangle = | \psi^{L}_{mn} \rangle \otimes | \psi^{R}_{mn} \rangle $ are the left and right Schmidt vectors and $\lambda_{mn} \geq 0$ is the entanglement spectrum of the $m$-th eigenvector of $H$. Now, inserting this expression into the diagonalization of $H$ \ref{Diag_form_H_II},
	\begin{align*}
		\hat{H}^{2} = \sum_{ m } \sum_{nn'} \Lambda^{2}_{m} \lambda_{mn} \lambda_{mn'} | \psi^{L}_{mn} ,\psi^{R}_{mn}  \rangle \langle \psi^{L}_{mn'} , \psi^{R}_{mn'} |
	\end{align*}
	and for the swapped $\hat{H}$,
	\begin{align*}
		\hat{H}_{S}^{2} =  \sum_{ m } \sum_{nn'} \Lambda^{2}_{m} \lambda_{mn} \lambda_{mn'} | \psi^{R}_{mn} ,\psi^{L}_{mn}  \rangle \langle \psi^{R}_{mn'} , \psi^{L}_{mn'} |
	\end{align*}
	Then, we have that
	\begin{align*}
		\text{Tr}[ \hat{H}_{S}^{2} \hat{H}^{2} ] =  \sum_{ij} \sum_{mm'nn'} \Lambda^{2}_{i}\Lambda^{2}_{j} \lambda_{in}\lambda_{in'} \lambda_{jm}\lambda_{jm'} \langle \psi^{R}_{jm} , \psi^{L}_{jm} | \psi^{L}_{in} , \psi^{R}_{in} \rangle \langle \psi^{L}_{in'} , \psi^{R}_{in'} | \psi^{R}_{jm'} , \psi^{L}_{jm'} \rangle
	\end{align*}
	This is equivalent to,
	\begin{align*}
		\text{Tr}[ \hat{H}_{S}^{2} \hat{H}^{2} ] =  \sum_{ij} \sum_{mm'nn'} \Lambda^{2}_{i}\Lambda^{2}_{j} \lambda_{in}\lambda_{in'} \lambda_{jm}\lambda_{jm'} \langle \psi^{R}_{jm}  | \psi^{L}_{in} \rangle \langle \psi^{L}_{jm} | \psi^{R}_{in} \rangle \langle \psi^{L}_{in'} | \psi^{R}_{jm'} \rangle \langle  \psi^{R}_{in'} | \psi^{L}_{jm'} \rangle
	\end{align*}
	For a phenomenological description of quantum states, this is highly desirable. Specifically, a variant of the Berry conjecture \cite{Stockmann1999Quantum} states that in a generic basis, large spectrum eigenvectors have coefficients that are drawn iid at random from a complex Gaussian (then normalized). If the Berry conjecture is true, the overlap between left and right Schmidt eigenvectors need to decay as 
	\begin{align}\label{Howell_Conjecture}
		| \langle \psi^{L}_{mk} | \psi^{R}_{m'k'} \rangle |^{2} \sim \mathcal{O}( \frac{1}{d} )
	\end{align}
	should be small for large $m>>1$ and $m'>>1$. The term $\text{Tr}[ \hat{H}^{\dagger}_{S}\hat{H}_{S}\hat{H}^{\dagger} \hat{H} ]$ suppresses matrices where this condition \ref{Howell_Conjecture} does not hold.

	\section{Conclusion}
	
	In this note, we have motivated a new class of random matrix models defined on $k$-fold tensor product spaces $W = V^{ \otimes k } = V \otimes V \otimes V \otimes ... \otimes V$. We show that $k$-fold invariance emerges naturally in two physical settings \ref{Appendix:Physical_Models}. Specifically, $k$-fold invariance is equivalent to the assumption that energy dependent only on the relative orientations of different subsystems. Thus, $k$-fold invariance is found in many models that `on average' have continuous symmetries. This is common in physical models as although isotropic disorder breaks continuous symmetries, the average over isotropic noise still retains this symmetry. The $k$-fold invariance assumption is also intimately related to the theory of gauge fields. Specifically, we have show that the unitary double model, which is a generalization of the quantum double model to the unitary group, satisfies the $k$-fold unitary invariance property. Using the Schur-Weyl decomposition, we have given a complete characterization of matrix distributions that are invariant under $k$-fold conjugation by the unitary, orthogonal and symplectic groups. The generalizations of the random matrix ensembles we have proposed seem extremely well suited for modeling Hamiltonians of quantum mechanical systems of many identical particles and may be of interest to both the quantum information and condensed matter community. Our work provides a representation theoretic method to build random matrix ensembles that display properties not observed in the standard Gaussian Ensembles.

	\emph{Future Work:}
	In this note, motivated a generalization of the standard random matrix ensembles. Specifically, we argue that there is a natural generalization of the standard Gaussian matrix ensembles \cite{Dyson_2004_Threefold} to ensembles that are invariant under $k$-fold local, as opposed to global, symmetry transformations. We only consider probability distributions with elements drawn from a Gaussian distribution. We can study non-Gaussian correlations by via diagrammatic methods. Ideally, we would like to perform a full renormalization group analysis of perturbations with $k$-fold symmetry. Furthermore, we would like to study properties of such ensembles in the limit $k \rightarrow \infty$. This is a difficult problem, but techniques in free probability \cite{Pappalardi_2022_Eigenstate,Fava_2023_designs} may allow for such asymptotic expansions.

	Disorder plays a crucial role in the physics of materials. Properties that are observable in real physical systems must be noise robust. For this reason, a full understanding of novel states of matter should requires random matrix theory. By constructing a natural generalization of the Gaussian ensembles, we believe that our work is a first step in developing random matrix models describing more exotic states of matter. We expect that the random matrix models proposed in this note would benefit from understanding gravity dual theories, similar to \cite{Jafferis2023MatrixModels}.

	\small
	\section*{Acknowledgments}
	Owen L. Howell thanks Shubhendu Trivedi for useful discussions. Owen L. Howell is grateful to the National Science Foundation Graduate Research Fellowship Program (NSF-GRFP) for continued financial support, without which this work would be impossible. Lastly, Michael Brodskiy and Owen L. Howell would like to thank the U.S. taxpayer, who is the unsung hero of modern science.
	\normalsize

	\bibliographystyle{plain} 
	\bibliography{refs} 
	\newpage
	\appendix

	\section{Representation Theory}
	Let $V$ be a vector space over the field $\mathbb{C}$. A representation $(\rho , V)$ of a group $G$ consists of $V$ and a group homomorphism $\rho : G \rightarrow \text{Hom}[V,V] $. By definition, the homomorphism $\rho$ satisfies
	\begin{align*}
		\forall g,g' \in G, \enspace \forall v \in V, \enspace \rho(g)\rho(g')v = \rho(gg') v 
	\end{align*}
	Heuristically, a group can be thought of as the embedding of an group (which is an abstract mathematical object) into a set of matrices. Two representation $(\rho,V)$ and $(\sigma,W)$ are said to be equivalent representations if there exists a matrix $\Phi$
	\begin{align*}
		\forall g\in G, \quad  \Phi \rho(g) = \sigma(g) \Phi
	\end{align*}
	The linear map $\Phi$ is said to be a $G$-intertwiner of the $(\rho,V)$ and $(\sigma , W)$ representations. The space of all $G$-intertwiners is denoted as $\Hom_{G}[ (\rho,V) , (\sigma , W) ]$. Specifically,
	\begin{align*}
		\Hom_{G}[ (\rho,V) , (\sigma , W) ] = \{ \enspace \Phi : V \rightarrow W \enspace | \enspace \forall g\in G, \enspace \Phi \rho(g) = \sigma(g) \Phi , \enspace \Phi \text{ is linear} \enspace \}
	\end{align*}
	The sum of two $G$-intertwiners is again $G$-intertwiner and $\Hom_{G}[ (\rho,V) , (\sigma , W) ]$ forms a vector space over $\mathbb{C}$. The vector space of of $G$-intertwiners from a representation to itself is called the $G$ endomorpism space of the representation $(\rho,V)$,
	\begin{align*}
		\text{End}_{G}[ ( \rho , V ) ] = \Hom_{G}[ (\rho,V) , (\rho ,V) ]
	\end{align*}
	which we will refer to as the endomorpism space of $(\rho,V)$. Much of classical group theory studies the structure of the intertwiners of representations \cite{Ceccherini_2008_Harmonic}. A representation $(\rho , V)$ is said to be a unitary representation if the vector space $V$ can be equipped with an inner product $\langle \cdot , \cdot \rangle$ such that
	\begin{align*}
		\forall g \in G, \enspace \forall v,w \in V, \quad \langle \rho(g) v , \rho(g) w \rangle = \langle  v ,  w \rangle
	\end{align*}
	The unitary theorem in representation theory \cite{Ceccherini_2008_Harmonic} says that any representation of a compact group $G$ is equivalent to a unitary representation of $G$. A representation is said to be reducible if it breaks into a direct sum of smaller representations. Specifically, a unitary representation $\rho$ is reducible if there exists an unitary matrix $U$ such that
	\begin{align*}
		\forall g\in G, \quad \rho(g) = U [ \bigoplus_{i=1}^{k} \sigma_{i}(g) ] U^{\dagger}
	\end{align*}
	where $k\geq2$ and $\sigma_{i}$ are smaller representations of $G$. The set of all non-equivalent unitary representations of a group $G$ will be denoted as $\hat{G}$. All representations of compact groups $G$ can be decomposed into direct sums of irreducible representations. Specifically, if $(\sigma , V)$ is a $G$-representation, 
	\begin{align*}
		(\sigma , V) = U[  \bigoplus_{ \rho \in \hat{G} } m^{\rho}_{\sigma}(\rho , V_{\rho} )  ] U^{\dagger}
	\end{align*}
	where $U$ is a unitary matrix and the integers $m^{\rho}_{\sigma}$ denote the number of copies of the irreducible $(\rho , V_{\rho} )$ in the representation $(\sigma , V)$.
	
	\section{Schur's Lemma}\label{Suppl_Schur_Lemma}
	
	Schur's lemma is one of the fundamental results in representation theory \cite{Zee_2016}. Let $G$ be a compact group. Let $(\rho , V)$ and $(\sigma , W)$ be irreducible representations of $G$. Then, 
	Schur's lemma states the following: Let $\Phi: V \rightarrow W$ be an intertwiner of  
	$(\rho , V)$ and $(\sigma , W)$. Then, $\Phi$ is either zero or the proportional to the identity map. In other words,
	\begin{align*}
		\text{ if } \forall g\in G, \enspace	\Phi  \rho(g) = \sigma(g) \Phi \implies \begin{cases}
			& \Phi \propto \mathbb{I} \text{ if }(\rho , V) = (\sigma , W)  \\
			& \Phi = 0 \text{ if else}
		\end{cases}
	\end{align*}
	Equivalently, if $(\rho , V)$ and $(\sigma , W)$ are irreducible representations, the space of intertwiners of representations satisfies
	\begin{align*}
		\Hom_{G}[ (\rho , V), (\sigma , W)  ] \cong 
		\begin{cases}
			& \mathbb{C} \text{ if } (\rho , V) = (\sigma , W)  \\
			& 0 \text{ if else}
		\end{cases}
	\end{align*}
	A corollary of Schur's lemma is the following: Let $(\rho , V)$ be a irreducible representation of $G$. Let $M \in \mathbb{C}^{d_{\rho}\times d_{\rho}}$ be a matrix. Suppose that 
	\begin{align*}
		\forall g \in G, \quad \rho(g)M = M\rho(g)
	\end{align*}
	holds. Then, $M$ is proportional to the identity matrix. The constant of proportionally can be determined by taking traces. Specifically,
	\begin{align*}
		M = \frac{\text{Tr}[M]}{d_{\rho}} \mathbb{I}_{d_{\rho}}
	\end{align*}
	
	\subsection{Extended Shur Lemma}\label{Suppl_Extended_Schur_Lemma}
	
	Schur's Lemma can be extended to reducible representations. Let $(\rho,V_{\rho})$ and $(\sigma, V_{\sigma})$ be $G$ representations which decompose into irriducibles as
	\begin{align*}
		(\rho,V_{\rho}) = U[\bigoplus_{ \tau \in \hat{G} } m^{\rho}_{\tau} (\tau , W_{\tau}   ) ]U^{\dagger} \quad (\sigma,V_{\sigma}) = V[ \bigoplus_{ \tau \in \hat{G} } m^{\sigma}_{\tau} (\tau , W_{\tau}   ) ] V^{\dagger}
	\end{align*}
	where $U, V$ are fixed unitary matrices that diagonalize the $\rho$ and $\sigma$ representations, respectively. Then, the vector space of intertwiners between $(\rho,V_{\rho})$ and $(\sigma, V_{\sigma})$ has dimension
	\begin{align*}
		\dim \Hom_{G}[ (\rho,V_{\rho}) , (\sigma,V_{\sigma}) ] = \sum_{ \tau \in \hat{G} }  m^{\rho}_{\tau}m^{\sigma}_{\tau}
	\end{align*}
	Furthermore, elements of the space $\Hom_{G}[ (\rho,V_{\rho}) , (\sigma,V_{\sigma}) ]$ have block structure. Specifically, any $\Phi \in \Hom_{G}[ (\rho,V_{\rho}) , (\sigma,V_{\sigma}) ]$ can be parameterized in block diagonal form as
	\begin{align*}
		\Phi = U[ \bigoplus_{ \tau \in \hat{G} } \Phi^{\tau} \otimes \mathbb{I}_{ d_{\tau} } ]V^{\dagger}
	\end{align*}
	and each block $\Phi^{\tau}$ is a $m^{\rho}_{\tau} \times m^{\sigma}_{\tau}$ matrix written as
	\begin{align*}
		\Phi^{\tau} = \begin{bmatrix}
			\Phi^{\tau}_{11}  & \Phi^{\tau}_{12} & ... & \Phi^{\tau}_{1m^{\sigma}_{\tau}}  \\
			\Phi^{\tau}_{21}  & \Phi^{\tau}_{22}  & ... & \Phi^{\tau}_{2m^{\sigma}_{\tau}}  \\
			... & ... & ... & ... \\
			\Phi^{\tau}_{m^{\rho}_{\tau}1} & \Phi^{\tau}_{m^{\rho}_{\tau}2}  & ... & \Phi^{\tau}_{m^{\rho}_{\tau}m^{\sigma}_{\tau}}  \\
		\end{bmatrix} 
	\end{align*}
	where each $\Phi^{\tau}_{ij} \in \mathbb{C}$ is a complex constant and $d_{\tau} = \dim (\tau , W_{\tau} )$ is the dimension of the irreducible $G$-representation $(\tau , W_{\tau} )$.
	
	\section{ Schur-Weyl Duality }\label{More_Schur-Weyl_Duality}

	Schur-Weyl Duality is a powerful tool in the representation theory of compact groups \cite{Weyl_1966_Classical}. In the literature there is some ambiguity as to the actual definition of what Schur-Weyl duality entails. Schur-Weyl Duality is sometimes referred to as the decomposition of the tensor products classical Lie groups. However, Schur-Weyl is actually a more general idea that can be used to decompose any $k$-fold tensor product of a representation of a compact group. Let $G$ be a compact group. Let $(\rho , V_{\rho} )$ be any representation of $G$. Consider the $k$-fold tensor product representation, $( \rho^{\otimes k} , V_{\rho}^{\otimes k} )$. This representation also forms a representation of the symmetric group of order $k$, as
	\begin{align*}
		\forall \sigma \in S_{k}, \enspace \forall g\in G, \quad S_{\sigma} \underbrace{[  \rho(g) \otimes \rho(g) \otimes ... \otimes \rho(g) ]}_{k-times} =\underbrace{[  \rho(g) \otimes \rho(g) \otimes ... \otimes \rho(g) ]}_{k-times}S_{\sigma} 
	\end{align*}
	so that the $G$ action and $S_{k}$ action are commutative.
	\begin{center}\label{Diagram:Completeness_Property}
		\begin{tikzcd}[row sep=large, column sep = large]\centering
			& V_{\rho}^{\otimes k}  \arrow{r}{ \rho^{\otimes k}(g) } \arrow{d}[swap]{ \hat{S}_{\sigma} } & V_{\rho}^{\otimes k}  \arrow{d}[swap]{ \hat{S}_{\sigma} } \\
			&  V_{\rho}^{\otimes k}  \arrow{r}{ \rho^{\otimes k}(g) } & V_{\rho}^{\otimes k} 
		\end{tikzcd}
	\begin{figure}[!h]
	\caption{`Square'-type commutative diagram for Schur-Weyl duality. The key observation in Schur-Weyl duality is that the $k$-fold tensor product action and the tensor permutation representation are commutative. This allows for definition of $G \times S_{k}$ action on the vector space $V_{\rho}^{\otimes k}$. Because of this, $( \Pi^{k}_{\rho} , V_{\rho}^{\otimes k} )$ forms a representation of the group $G \times S_{k}$.  }
	\end{figure}
	\end{center}

	\noindent
	Let us define the action $ \Pi^{k}_{\rho} $ on the vector space $V_{\rho}^{\otimes k}$ as the following
	\begin{align*}
		\forall g\in G, \enspace \forall \sigma \in S_{k}, \enspace \forall w_{i_{1}i_{2}...i_{k}} \in V_{\rho}^{\otimes k}, \quad \Pi^{k}_{\rho}( g , \sigma ) w_{i_{1}i_{2}...i_{k}} = \sum_{j_{1}=1}^{d}\sum_{j_{2}=1}^{d}...\sum_{j_{k}=1}^{d}  \rho(g)_{i_{\sigma(1)}j_{1}}\rho(g)_{i_{\sigma(2)}j_{2}}...\rho(g)_{i_{\sigma(k)}j_{k}} w_{j_{1}j_{2}...j_{k}}
	\end{align*}
	Note that this action is well defined and can be performed by matrix multiplication followed by permutation or permutation followed by matrix multiplication. For this reason, $( \Pi^{k}_{\rho} , V^{\otimes k} )$ is a well defined representation of the group $G \times S_{k}$. The representation $( \Pi^{k}_{\rho} , V_{\rho}^{\otimes k} )$ is in general not reducible and will decompose into irreducible representations of $ G \times S_{k}$. Irreducible representations of $G \times S_{k}$ are tensor products of irreducible representations of $G$ and irreducible representations of $S_{k}$. Thus, we have the following decomposition,
	\begin{align*}
		( \Pi^{k}_{\rho} , V^{\otimes k}_{\rho} )  \cong \bigoplus_{ \tau \in \hat{G} } \bigoplus_{ \lambda \vdash k } m^{ k \tau \lambda }_{ \rho }  (\tau, V_{\tau} ) \otimes ( \lambda , V_{\lambda}) 
	\end{align*}
	where $m^{ k \tau \lambda }_{ \rho }$ are integers counting the number of copies of the $(\tau, V_{\tau} ) \otimes ( \lambda , V_{\lambda})$ irreducible in $( \Pi^{k}_{\rho} , V_{\rho}^{\otimes k} )$. Thus, the tensor product space decomposes into vector subspaces that are characterized by their transformation properties based on $G$ action and tensor index permutations.

	\subsection{Unitary Schur-Weyl Duality}
	
	Let us apply the more general Schur-Weyl formalism to the case of the unitary group $U(d)$. Irreducible representations of $U(d)$ are countably infinite and are in one-to-one correspondence with integer partitions \cite{Zee_2016,Weyl_1966_Classical}. Let $\lambda = ( \lambda_{1} , \lambda_{2} , ... , \lambda_{m} )$ be a partition with $\lambda_{1} \geq \lambda_{2} \geq ... \geq \lambda_{m}$. The irreducible representation of $U(d)$ associated to the partition $\lambda$ will be denoted as $(U_{\lambda} , V_{\lambda} )$. Let $( U_{1} , \mathbb{C}^{d} )$ be the fundamental $d$-dimensional representation of $U(d)$ defined as the $\lambda = (1)$ partition,
	\begin{align*}
		U_{d} = \{ \enspace U \enspace | \enspace U^{\dagger}U = \mathbb{I}_{d} = UU^{\dagger} \enspace \}
	\end{align*}
	Consider the $k$-fold tensor product decomposition,
	\begin{align*}
		( \mathbb{C}^{d} )^{\otimes k}  = \bigoplus_{ \lambda \vdash (k,d) } V_{\lambda} \otimes \lambda
	\end{align*}
	where $\lambda \vdash (k,d) $ denotes partitions of the integer $k$ with no more than $d$ summands, i.e.
	\begin{align*}
		\lambda \vdash (k,d) \implies \lambda = (\lambda_{1}, \lambda_{2}, ... , \lambda_{m} ), \enspace \text{ s.t } \lambda_{1} \geq \lambda_{2} \geq ... \geq \lambda_{m} \text{ s.t. } \sum \lambda_{i} = k \text{ and } m\leq d
	\end{align*}
	A celebrated theorem of Weyl \cite{Weyl_1966_Classical} states that the representations $( U_{\lambda} , V_{ \lambda } )$ exhaust all representations of the $d$-dimensional unitary group $U(d)$. 
	
	\subsubsection{ Unitary Group Tensor Product Rules }
	
	For a complete discussion of diagrammatic methods for computing tensor products of irreducible representations of the unitary group, please see \cite{DiFrancesco1997conformal}. We will be interested in tensor products of irreducible representations of $U(d)$. Let $\lambda$ and $\lambda'$ be two partitions. Let $V_{\lambda}$ and $V_{\lambda'}$ be the corresponding irreducible representations of $U(d)$. Then, consider the tensor product
	\begin{align*}
		V_{ \lambda } \otimes V_{ \lambda' } \cong  \bigoplus_{ n=1 }^{\infty} \bigoplus_{ \mu \vdash n } m^{\mu}_{ \lambda \lambda' } V_{ \mu } 
	\end{align*}
	so that the index $\mu$ ranges over all integer partitions and $m^{\mu}_{ \lambda \lambda' }$ are integers that count the muplicity of the irreducible representation $V_{ \mu }$ in the tensor product $V_{\lambda} \otimes V_{\lambda'}$. Using Schur-Weyl Duality, we can derive an exact expression for tensor product rules $m^{\mu}_{\lambda \lambda'}$ of the unitary group in terms of the branching rules of the symmetric group. To begin, consider the trivial relation
	\begin{align*}
		( \mathbb{C}^{d} )^{\otimes k} \otimes ( \mathbb{C}^{d} )^{\otimes k'} = ( \mathbb{C}^{d} )^{\otimes ( k + k' ) }  
	\end{align*}
	for any integers $k$ and $k'$. Then, using the vector space decomposition in the Schur-Weyl duality, we have an isomorphism of vector spaces
	\begin{align*}
		\underbrace{[ \bigoplus_{ \lambda \vdash (k,d) } V_{\lambda} \otimes \lambda ]}_{ ( \mathbb{C}^{d} )^{k} } \otimes \underbrace{ [ \bigoplus_{ \lambda' \vdash (k',d) } V_{\lambda'} \otimes \lambda'  ] }_{ ( \mathbb{C}^{d} )^{k'} } \cong \underbrace{ [ \bigoplus_{ \mu \vdash (k + k',d) } V_{\mu} \otimes \mu  ] }_{ (\mathbb{C}^{d})^{k+k'} }
	\end{align*}
	This is a representation of the group $U(d)\times S_{k} \times S_{k'}$. Expanding out the tensor product of the left hand side, we have that
	\begin{align*}
		\bigoplus_{ \lambda \vdash (k,d) } \bigoplus_{  \lambda' \vdash (k',d) } [V_{\lambda} \otimes V_{\lambda'} ] \otimes (\lambda \otimes \lambda') = \bigoplus_{ \mu } \bigoplus_{ \lambda \vdash (k,d) } \bigoplus_{ \lambda' \vdash (k',d) } m^{\mu}_{ \lambda \lambda' }V_{\mu} \otimes (\lambda \otimes \lambda')
	\end{align*}
	Now, consider the group restriction of the left side from $S_{k+k'}$ to the subgroup $ S_{k} \times S_{k'} \subseteq S_{k+k'}$. Let $\mu \vdash (k+k')$ be a irreducible representation of $S_{k+k'}$. Under the group restriction
	\begin{align*}
		\Res^{ S_{k+k'} }_{ S_{k} \times S_{k'} }[  \mu  ] = \bigoplus_{ \lambda \vdash k } \bigoplus_{ \lambda' \vdash k' } B^{ \lambda \lambda' }_{\mu} ( \lambda \otimes \lambda' ) 
	\end{align*}
	where $B^{ \lambda \lambda' }_{\mu}$ are the branching rules which count how many copies of the irreducible $\lambda \otimes \lambda'$ are contained in the restriction of $\mu$. Branching rules for the symmetric group have been thoroughly studied \cite{james1981representation}. Under group restriction from $S_{k+k'} \rightarrow S_{k} \times S_{k'}$, the isomorphism of vector spaces becomes an isomorphism of group representations. Under restriction
	\begin{align*}
		\bigoplus_{ \mu \vdash (k+k',d) } V_{\mu} \otimes \mu \rightarrow \bigoplus_{ \mu \vdash (k+k',d) } \bigoplus_{ \tau \vdash k } \bigoplus_{ \tau' \vdash k' } B^{\mu}_{\tau \tau'} V_{\mu} \otimes [  \tau \otimes \tau' ]
	\end{align*}
	Two representations are equivalent if and only if they have identical decomposition of irriducibles. This relation can only hold if, for any $\lambda \vdash k$ and $\lambda' \vdash k'$ the relation
	\begin{align*}
		V_{\lambda} \otimes V_{\lambda'} = \bigoplus_{ \mu \vdash (k+k') } B^{ \mu }_{ \lambda \lambda' } V_{ \mu }
	\end{align*}
	holds. Thus, the tensor product of the $\lambda$ and $\lambda'$ irreducibles of $U(d)$ are completely determined by the branching rules of irreducible representations of the symmetric group. Branching rules of the symmetric group have been thoroughly studied in representation theory \cite{james1981representation}. When $k = k'$ there is additional simplification due to exchange. Specifically, consider the decomposition
	\begin{align*}
		( \mathbb{C}^{d} )^{\otimes k } \otimes ( \mathbb{C}^{d} )^{\otimes k } \cong ( \mathbb{C}^{d} )^{\otimes 2k }
	\end{align*}
	The left hand side of this expression has a $\mathbb{Z}_{2}$-action. Specifically, define the operator $\hat{S}$ as swapping the left and right copy of $( \mathbb{C}^{d} )^{\otimes k }$.

	
	

	\section{Lie Group Theory}
	
	Lie group theory is the study of continuous groups. We review some basic concepts of Lie group theory. A full treatment of Lie group theory can be found in \cite{Zee_2016,Hall_2015_Lie,Weyl_1966_Classical,DiFrancesco1997conformal}. A Lie group $G$ is a group that is also a smooth manifold with the requirement that, for all $g,h\in G$, the map $g \times h \rightarrow gh : G \times G \rightarrow G$ is smooth and the map $g \rightarrow g^{-1} : G \rightarrow G$ is smooth. A homeomorphism of Lie groups is a smooth map $\Phi : G \rightarrow H$ that satisfies the relation
	\begin{align*}
		\forall g,g' \in G, \quad \Phi(gg') = \Phi(g) \Phi(g')
	\end{align*}
	Representations of Lie groups are defined in the same way as representations of finite groups. Let $V$ be a vector space. A representation of a Lie group is a Lie group homeomorpism $\rho : G \rightarrow GL(V)$ and a vector space $V$ satisfying,
	\begin{align*}
		\forall g \in G, \enspace \forall v \in V, \quad \rho(gg') v = \rho(g)\rho(g') v
	\end{align*}
	
	\subsection{Lie Algebra}
	A Lie algebra $\frak{g}$ is a vector space equipped with a anti-symmteric two-form $[ \cdot, \cdot] : \frak{g} \times \frak{g} \rightarrow \frak{g}$ which satisfies the Jacobi identity,
	\begin{align*}
		\text{Jacobi: } [X,[Y,Z]] + [Y,[Z,X]] + [Z,[X,Y]] = 0
	\end{align*}
	A homeomorphism of Lie algebras is a map $\phi : \frak{g} \rightarrow \frak{h}$ that preserves the Lie bracket of $\frak{g}$ so that
	\begin{align*}
		\forall X,Y \in \frak{g}, \quad  \phi( [ X , Y ] ) = [ \phi(X) , \phi(Y) ]
	\end{align*}
	Let $X_{i}$ be a basis of the Lie algebra $\frak{g}$. The Lie algebra $\frak{g}$ is called semi-simple if there is no proper subset $J_{i}$ of the $X_{i}$ such that the $J_{i}$ are an idea of $\frak{g}$ under the Lie bracket operator $[ \cdot , \cdot ]$. Let $X_{i}$ be a basis of the Lie algebra $\frak{g}$. The structure constants $f^{k}_{ij}$ of $\frak{g}$ are defined as
	\begin{align*}
		[X_{i},X_{j}] = \sum_{j} f^{k}_{ij}X_{k}
	\end{align*}
	so that the constants $f^{k}_{ij}$ are the decomposition of the Lie bracket in the vector space $\frak{g}$. Let $V$ be a vector space. We can similarly speak of a Lie algebra representation as a homeopmorpism $\sigma: \frak{g} \rightarrow GL(V)$ that preserves Lie bracket structure
	\begin{align*}
		\forall X , Y \in \frak{g}, \quad \sigma( [X,Y] ) = [ \sigma(X) , \sigma(Y) ]
	\end{align*}
	If $G$ is a connected group, the map $\exp : \frak{g} \rightarrow G$, is defined as
	\begin{align*}
		\forall X \in \frak{g}, \quad \exp(itX) = \sum_{n=0}^{\infty} \frac{(it)^{n}}{n!} X^{n}
	\end{align*}
	The key property of $\exp$ is that the exponential map $\exp$ commutes with homeomorphism of algebra and group \ref{Diagram:Exponential Map}, so that there is an isomerism between Lie algebra representations and Lie group representations.

	\begin{center}\label{Diagram:Exponential Map}
		\begin{tikzcd}[row sep=large, column sep = large]\centering
			& \frak{g}  \arrow{r}{ d\Phi|_{e} } \arrow{d}[swap]{ \exp } & \frak{h} \arrow{d}[swap]{ \exp } \\
			&  G  \arrow{r}{ \Phi } & H
		\end{tikzcd}
		\begin{figure}[h!]
			\caption{ The exponential map: Let $\Phi: G \rightarrow H$ be a homeomorpism of groups. Let $d\Phi |_{e}: \frak{g} \rightarrow \frak{h}$ be the derivative map evaluated at the identity of $G$. Then, the above map is commutative.  }
		\end{figure}
	\end{center}

	\subsection{Adjoint Representation}
	\noindent
	The adjoint (sometime called the little adjoint) $\text{ad}$ representation is a canonical representation of a Lie algebra. The adjoint action is defined as
	\begin{align*}
		\text{ad}(X)Y = [X,Y]
	\end{align*}
	The adjoint action satisfies
	\begin{align*}
		[\text{ad}(X),\text{ad}(Y)] = \text{ad}( [X,Y] )
	\end{align*}
	which preserves the Lie bracket structure and is a valid Lie algebra representation. The adjoint action acts directly on $\hat{g}$, and the dimension of the adjoint representation is the dimension of the vector space $\hat{g}$.
	
	There is an analogous adjoint (sometimes called the big adjoint) $\text{Ad}$  representation of the Lie group $G$ on $\frak{g}$. Consider the conjugation map $\Phi_{g}: G \rightarrow G$ on the Lie group $G$ given by
	\begin{align*}
		\Phi_{g}(h) = ghg^{-1}
	\end{align*}
	the conjugation map is an Lie automorpism of $G$. The adjoint map $Ad_{g}$ evaluated at $g\in G$ is then the conjugation map evaluated at the identity
	\begin{align*}
		\forall g\in G, \quad Ad_{g} =  d \Phi_{g} |_{e} : T_{e}(G) \rightarrow T_{e}(G)
	\end{align*}
	so that for fixed $g\in G$, $Ad_{g} : \frak{g} \rightarrow \frak{g}$. Thus, $Ad_{g} : G \rightarrow \text{aut}( \frak{g} ) $. Let $X \in \frak{g}$,
	\begin{align*}
		\forall g\in G, \quad Ad_{g} X = \frac{d}{dt}[g \exp(tX) g^{-1} ]|_{t = 0}
	\end{align*}
	Note that
	\begin{align*}
		\forall g,g' \in G, \quad Ad_{g}\circ Ad_{g'} = Ad_{gg'}
	\end{align*}
	so that $(Ad , \frak{g} )$ is a Lie group representation of $G$ with dimension equal to the vector space dimension of $\frak{g}$. Let $\langle \cdot , \cdot \rangle$ be an inner product on $\frak{g}$. The inner product $\langle \cdot , \cdot \rangle$ is said to be $Ad$-invariant if and only if,
	\begin{align*}
		\forall g\in G, \enspace \forall x,y \in \frak{g}, \quad \langle x , y \rangle = \langle Ad_{g}x , Ad_{g} y \rangle
	\end{align*}
	
	\subsection{Killing Form}
	
	\noindent
	The Killing form $K$ is a symmetric bi-linear form on a Lie algebra $\frak{g}$. Specifically, $K$ is defined as
	\begin{align*}
		K(X,Y) = \text{Tr}[  \text{ad}(X)\text{ad}(Y) ]
	\end{align*}
	Using the cyclic properties of the trace,
	\begin{align}\label{Killing Trace}
		K( X , [Z,Y] ) + K( [Z,X] , Y) = 0 
	\end{align}
	The Killing form is essentially unique. It is (up to multiplication) the only inner product satisfying the property \ref{Killing Trace}.
	
	The Killing form can be written in terms of the structure constants $f^{k}_{ij}$ as
	\begin{align*}
		K( A^{i} X_{i}, B^{j}X_{j} ) = \sum_{k} f^{k}_{ij}f_{ji}^{k} A^{i}B^{j}
	\end{align*}
	So that as an element of $\frak{g}^{\star} \otimes \frak{g}^{\star}$ the Killing form is given by
	\begin{align*}
		K = \sum_{km=1} f^{k}_{i m}f^{m}_{ j k} e^{i} \otimes e^{j}
	\end{align*}
	where $ \frak{g}^{\star} = \text{span}[ e^{i} ]_{i=1}^{r} $ is the dual space of $\frak{g}$. Importantly, the Killing form is an $Ad$-invariant inner product,
	\begin{align*}
		\forall g\in G, \quad K(X,Y) = K( \text{Ad}_{g}X , \text{Ad}_{g}Y )
	\end{align*}

	
	\subsubsection{Cartan Sub-Algebra}
	
	A Cartan sub-algebra $\frak{h}$ is a maximal commuting set of elements of $\frak{g}$. A Cartan sub-algebra is closed under commutation and satisfies
	\begin{align*}
		\forall x,y\in \frak{h}, \quad [ x ,y ] = 0 
	\end{align*}
	The dimensions of $\dim \frak{h} = r$ is called the rank of $\frak{g}$. Let $ \{ h^{i} \}_{i=1}^{r}$ be a basis of $\frak{h}$. The remaining elements of $\frak{g}$ will be denoted as $E^{\alpha}$ where
	\begin{align*}
		\forall h \in \frak{h}, \quad [  h^{i} , E^{\alpha} ] =  \alpha^{i} E^{\alpha}
	\end{align*}
	so that the $E^{\alpha}$ are eigenvectors of the $h^{i}$ operators. The vectors $\alpha = ( \alpha^{1} , \alpha^{2}, ..., \alpha^{r} )$ are called roots. The operator $E^{\alpha}$ is called the ladder operator associated to the root $\alpha$. Let $\Phi$ denote all the roots of $\frak{g}$. The Lie algebra $\frak{g}$ then decomposes as a direct sum of the Cartan sub-algebra and the roots
	\begin{align*}
		\frak{g} = \frak{h} \bigoplus_{ \alpha \in \Phi } E^{\alpha}
	\end{align*}
	Root systems have a reflection symmetry. Specifically, if $\alpha$ is a root, then $-\alpha$ is also a root as
	\begin{align*}
		[h^{i} , E^{\alpha} ] = \alpha^{i} E^{\alpha} \implies [h^{i} , (E^{\alpha})^{\dagger} ] = -\alpha^{i} (E^{\alpha})^{\dagger} 
	\end{align*}
	Using the Jacobi Identity, we have that
	\begin{align*}
		\forall h \in \frak{h}, \quad [  h^{i} , [E^{\alpha}, E^{\beta}] ] =  (\alpha + \beta )^{i} E^{\alpha + \beta}
	\end{align*}
	thus, the commutator of two roots satisfies
	\begin{align*}
		& [E^{\alpha}, E^{\beta}] = N_{\alpha , \beta } E^{\alpha + \beta } \text{ if } \alpha \ne -\beta \\
		& [E^{\alpha} , E^{-\alpha}] = \sum_{i=1}^{r} C_{i}(\alpha) h^{i}
	\end{align*}
	where $N_{\alpha,\beta}$ and $C_{i}(\alpha)$ are constants. The constant $C_{i}(\alpha)$ can be determined using the Jacobi relation. We have that
	\begin{align*}
		[ h^{i} , [E^{\alpha},E^{-\alpha}] ] + [ E^{\alpha} , [E^{-\alpha} , h^{i}] ] + [ E^{-\alpha} , [h^{i} , E^{\alpha}] ] = 0
	\end{align*}
	Using the definition of roots, we have that
	\begin{align*}
		[ h^{i} , [E^{\alpha},E^{-\alpha}] ] + 2\alpha^{i}[ E^{\alpha} , E^{-\alpha} ]  = 0
	\end{align*}
	Thus, $[E^{\alpha},E^{-\alpha}]$ must be given by
	\begin{align*}
		[ E^{\alpha},E^{-\alpha} ] = C( \alpha ) \sum_{i=1}^{r} \alpha^{i}h^{i}
	\end{align*}
	The root $\alpha(h) : \frak{h} \rightarrow \mathbb{C}$ is the eigenvector of $x$ in $[h,\cdot]$. Note that each root $\alpha : \frak{h} \rightarrow \mathbb{C}$ can be viewed as an element of the dual space $\frak{h}^{\star}$ of $\frak{h}$. An orientation on a root system $\alpha$ is a choice of roots $\Phi^{+} \subset \Phi$ such that either $\alpha$ or $-\alpha$ is contained in $\Phi^{+}$, but not both. If the Lie algebra $\frak{g}$ has an inner product, we can identify $\frak{h}^{\star}$ with $\frak{h}$. We can identify the dual $\frak{h}^{\star}$ with $\frak{h}$ via the canonical isomorphism $J : \frak{h}^{\star} \rightarrow \frak{h}$
	\begin{align*}
		J[x](y) = K( x , y )
	\end{align*}
	where $K( \cdot , \cdot )$ is the Killing form on $\frak{g}$. The Killing form induces a inner product on the root space. Let $\alpha$ and $\beta$ be roots. We can then define the inner product on roots
	\begin{align*}
		( \alpha , \beta ) = K( \sum_{i=1}^{r} \alpha^{i}h^{i} , \sum_{i=1}^{r} \beta^{i}h^{i}  )  = \sum_{i=1}^{r} \alpha^{i}\beta^{i}
	\end{align*}
	The Killing form then defines a inner product in the dual space $\frak{h}^{\star}$ via
	\begin{align*}
		( \alpha , \beta ) = K( \alpha \cdot h , \beta \cdot h )
	\end{align*}

	\subsection{Weights}
	
	A weight vector $\lambda = (\lambda^{1} , \lambda^{2}, ... , \lambda^{r} )$ is a basis such that
	\begin{align*}
		\forall h^{i}, \quad h^{i} | \lambda \rangle = \lambda^{i} | \lambda \rangle
	\end{align*}
	Using the commutation relations $[h^{i},E^{\alpha}] = \alpha^{i} E^{\alpha}$, we have that
	\begin{align*}
		h^{i}[ E^{\alpha}| \lambda \rangle ] = (\lambda^{i} + \alpha^{i} ) [ E^{\alpha} | \lambda \rangle ]
	\end{align*}
	so that the operator $E^{\alpha}$ shifts the weight vector $\lambda$,
	\begin{align*}
		E^{\alpha}|\lambda \rangle \propto | \lambda + \alpha \rangle
	\end{align*}
	The operator $E^{\alpha}$ is said to terminate the weight vector $\lambda$ is there exists an integer $p \in \mathbb{Z}$ such that
	\begin{align*}
		( E^{\alpha} )^{p} | \lambda \rangle = 0
	\end{align*}
	For finite representations, all the root operators $E^{\alpha}$ must terminate each weight vector $| \lambda \rangle $. Thus, we must have that
	\begin{align*}
		\frac{2(\alpha , \lambda)}{| \alpha |^{2}} \in \mathbb{Z}
	\end{align*}
	This is called the Cartan relation. The Cartan relation forces the root and weight space to satisfy a set of natural geometric relations, allowing for a complete classification of simple Lie algebras.

	\subsection{Structures of Root Systems}
	
	The rank of the Cartan sub-algebra $\frak{h}$ is in general much less than the dimension of the full Lie algebra $\frak{g}$. Let $\{ \beta_{i} \}_{i=1}^{r}$ be a basis of $\frak{h}^{\star}$. Then, any root may be expanded as
	\begin{align*}
		\forall \alpha \in \Phi, \quad \alpha = \sum_{i=1}^{r} n_{i} \beta_{i}
	\end{align*}
	where $n_{i}$ are integers. Roots with the first non-zero $n_{i}>0$ are called positive roots and denoted as $\Phi_{+}$. A simple root is a root that cannot be written as the sum of two positive roots. The set of simple roots is denoted as $\Delta$. There are exactly $r$ simple roots. For any two simple roots, we define the Cartan matrix
	\begin{align*}
		\alpha_{i}, \alpha_{j} \in \Delta, \quad A_{ij} = \frac{2\langle \alpha_{i}, \alpha_{j} \rangle}{|\alpha_{j}|^{2}}
	\end{align*}
	To each root $\alpha \in \Phi$, we associate a dual root $\alpha^{\wedge}$, defined as
	\begin{align*}
		\alpha^{\wedge} = \frac{2\alpha}{|\alpha|^{2}}
	\end{align*}
	Using this definition, the Cartan matrix can be written as
	\begin{align*}
		A_{ij} = \langle \alpha_{i} , \alpha^{\wedge}_{j} \rangle
	\end{align*}

	\subsubsection{Fundamental Weights}
	
	\noindent
	The fundamental weights are defined as the normalized coroots with
	\begin{align*}
		( \omega_{i} , \alpha^{\wedge}_{j} ) = \delta_{ij}
	\end{align*}
	Any weight vector can be expanded in the fundamental weight basis as
	\begin{align*}
		\lambda = \sum_{i=1}^{r} \lambda_{i} \omega_{i}
	\end{align*}
	where $\lambda_{i} = ( \lambda , \alpha^{\wedge}_{i} )$ are called the Dynkin labels of $\lambda$. The Weyl vector $\rho$ is defined as the sum of all fundamental weights
	\begin{align*}
		\rho = \sum_{i=1}^{r} \omega_{i}
	\end{align*}
	

	\subsection{Weyl Group}
	Consider the hyperplane defined by the equation
	\begin{align*}
		H_{\alpha} = \{ \enspace h \enspace | \enspace \langle \alpha , h \rangle > 0 \enspace \}
	\end{align*}
	For any root $\alpha \in \Phi$, we can reflect around the hyperplane defined by $H_{\alpha}$. The set of all reflections forms a group. Which is called the Weyl group $W$. Specifically, for any two roots $\beta$ and $\alpha$, the Weyl reflection of $\beta$ with respect to $\alpha$ is given by
	\begin{align*}
		s_{\alpha} \beta = \beta - ( \alpha^{\wedge} , \beta ) \alpha
	\end{align*}
	Because roots and weights live in the same space, the Weyl group also acts on weight vectors $|\lambda \rangle$ via
	\begin{align*}
		s_{\alpha} | \lambda \rangle = | \lambda \rangle - ( \alpha^{\wedge} , \lambda ) | \alpha \rangle
	\end{align*}
	The Weyl group action on both weights and roots is unitary,
	\begin{align*}
		\text{Roots: } \forall w \in W, \enspace \forall \alpha,\alpha' \in \Phi \quad (  \alpha , \alpha' ) = (  w \alpha , w \alpha' ) \\
		\text{Weights: }\forall w \in W, \quad (  \lambda , \lambda' ) = (  w \lambda , w \lambda' )
	\end{align*}
	It will be useful to define the Fredenhall operator $D_{\rho}$ as 
	\begin{align*}
		D_{\rho} = \prod_{ \alpha \in \Phi^{+} } (  \exp(\alpha/2) - \exp(-\alpha/2 ) )
	\end{align*}
	using the definition of the Weyl group, this can be written in terms of the Weyl vector as
	\begin{align*}
		D_{\rho} = \sum_{w \in W} \eta( w ) \exp( w \rho )
	\end{align*}
	where $\eta(w): W \rightarrow \pm 1$ is the sign function of $W$.
	\subsection{ Weyl Chamber}
	The action of the Weyl group $W$ on the root space splits the root space into $|W|$ isomorphic subspaces called chambers. The Weyl chamber defined as 
	\begin{align*}
		W_{c} = \{ \enspace \lambda \enspace | \enspace \forall w\in W, \enspace \forall \alpha_{i} \in \Delta \quad ( w\lambda , \alpha_{i} ) \geq 0  \enspace \}
	\end{align*}
	The discriminant function $\delta_{ \frak{g} }(x): \frak{h} \rightarrow \mathbb{C}$ is defined as
	\begin{align*}
		\forall x \in \frak{h}, \quad   \delta_{ \frak{g} }(x) = \prod_{ \alpha \in \Phi^{+} } \langle \alpha , x \rangle
	\end{align*}
	which is the products of the inner product of the Cartan element$x \in \frak{h}$ with all positive roots.

	\subsection{Highest Weight Representations}
	
	A highest weight vector $| \lambda \rangle$ is a weight that is decimated by each positive root,
	\begin{align*}
		\forall \alpha \in \Phi^{+}, \quad E^{\alpha} | \lambda \rangle = 0
	\end{align*}
	There is a bijection between highest weight representations and irreducible Lie algebra representations. Specifically, from a highest weight vector $| \lambda \rangle$, we can form the descendent states
	\begin{align*}
		\forall \alpha_{i} \in \Phi^{+}, \quad E^{-\alpha_{1}} E^{-\alpha_{2}} ... E^{-\alpha_{m}} | \lambda \rangle
	\end{align*}
	Descent states form representations of the Lie algebra $\frak{g}$. The set of all descendent states of the highest weight vector $|\lambda \rangle$ is denoted as $L_{\lambda}$.
	
	The descendent states $L_{\lambda}$ generate representation of the Lie algebra $G$. Specifically,
	\begin{align*}
		&\text{Cartan Subgroup: } \exp( \sum_{i=1}^{r} \theta_{i} h^{i} ) | \lambda \rangle = \exp( \sum_{i=1}^{r} \theta_{i} \lambda^{i} ) | \lambda' \rangle \\
		& \text{ Lie Algebra: } \exp( t E^{\alpha} ) | \lambda' \rangle \in L_{\lambda}
	\end{align*}
	Thus, highest weight states generate representations of Lie groups. However, we have to keep track of both the multiplicities of the states in $L_{\lambda}$ and be able to generate a basis for $L_{\lambda}$. Define the formal exponential $\exp( \mu )$ as a placeholder, where for all weights $\lambda$ and $\lambda'$,
	\begin{align*}
		& \exp( \lambda + \lambda' ) = \exp( \lambda ) \exp( \lambda' ) \\
		& \exp( \lambda )( \lambda' ) = \exp( (\lambda , \lambda' ) )
	\end{align*}
	The character of the highest weight representation $|\lambda \rangle$ is then defined as
	\begin{align*}
		\chi_{\lambda} =  \sum_{ \lambda' \in L_{\lambda} } \text{Mult}_{\lambda}[ \lambda' ] \exp(\lambda')
	\end{align*}
	where the integer $\text{Mult}_{\lambda}[ \lambda' ]$ counties the number of copies of the descendent state $|\lambda' \rangle$ in the $|\lambda \rangle$ highest weight representation. In general, calculating the Lie algebra characters is difficult. However, it can be show that the Freudenthal operator satisfies
	\begin{align*}
		D_{\rho} \chi_{\lambda}  = D_{\rho + \lambda}
	\end{align*}
	Thus, we have that
	\begin{align}\label{Weyl Charecter Formula}
		\chi_{\lambda} = \frac{D_{\rho + \lambda}}{D_{\rho}} 
	\end{align}
	This \ref{Weyl Charecter Formula} is called the Weyl character formula. Using \ref{Weyl Charecter Formula}, the dimension of a highest weight representation $|\lambda \rangle$ is given by
	\begin{align*}
		d_{\lambda} = \dim \lambda = \prod_{ \alpha \in \Phi^{+} } \frac{(\rho + \lambda , \alpha)}{(\rho , \alpha)}
	\end{align*}

	
	
	

	\section{ Harish-Chandra Integral Formula }
	
	The Harish-Chandra integrals were discovered by Harish-Chandra in his development of the theory of harmonic analysis on semi-simple Lie groups. The HCIZ integrals \cite{Itzykson1979Planar} are a special case of the more general Harish-Chandra formula. Let $G$ be a semi-simple group. Let $\text{Ad} : G \rightarrow \text{Aut}( \frak{g} )$ be the adjoint operator on $G$. Let $W$ be the Weyl group of $G$. Let $ \langle  \cdot , \cdot \rangle : \frak{g} \times \frak{g} \rightarrow \mathbb{C} $ be any $\text{Ad}$-invariant inner product on $\frak{g}$. Then, the Harish-Chandra formula evaluates integrals of the form
	\begin{align*}
		\int_{ g \in G }dg \text{ } \exp( \langle  \text{Ad}_{g}(x) , y \rangle )
	\end{align*}
	in terms of summations over the the Weyl group $W$. Specifically, 
	\begin{align*}
		\int_{ g \in G }dg \text{ } \exp( \langle \text{Ad}_{g}(x) , y \rangle ) = \frac{1}{ \text{Vol}(W) } \sum_{ w \in W } \text{sign}(w) \exp( \langle w(x) , y \rangle )
	\end{align*}
	where $w(x)$ is the lattice vector of $x$ on $W$ and $\text{sign} : W \rightarrow \pm 1$ is the sign function.

	\section{Representation Theory of Unitary Group $U(d)$ }
	
	The representation theory of the group $U(d)$ was worked out in the early 1900s by Jacobi, Schur and Weyl, among others. The representation theory of the group $U(d)$ is especially elegant and is intimately related to the representation theory of the symmetric group. The unitary group $U(d)$ is both semi-simple and compact so the set of irreducible representations of $U(d)$ are countably infinite. Let $\lambda = ( \lambda_{1} , \lambda_{2} , ... , \lambda_{m} )$ be an integer partition with $\lambda_{1} \geq \lambda_{2} \geq ... \geq \lambda_{m}$. Characters of irreducible representations are given by
	\begin{align*}
		s_{\lambda}( z_{1} , z_{2} , ... , z_{m}  ) = \chi_{\lambda}(z) : ( \mathbb{C}^{\times} )^{m} \rightarrow \mathbb{C}
	\end{align*}
	where the $s_{\lambda}$ are called called Schur functions. Define the function
	\begin{align*}
		a_{ \lambda_{1} , \lambda_{2} , ... , \lambda_{m} }(z_{1},z_{2},...,z_{m} ) = \det \begin{bmatrix}
			z^{ \lambda_{1} + m - 1}_{1} & z^{ \lambda_{1} + m - 1}_{2} & ... & z^{ \lambda_{1} + m - 1}_{n} \\
			z^{ \lambda_{2} + m - 2}_{1} & z^{ \lambda_{2} + m - 2}_{2} & ... & z^{ \lambda_{2} + m - 2}_{n} \\
			...  & ... & ... & ... \\
			z^{ \lambda_{n} }_{1} & z^{ \lambda_{n} }_{2} & ... & z^{ \lambda_{n} }_{n} \\
		\end{bmatrix}
	\end{align*}
	The Schur function is the defined by
	\begin{align*}
		s_{\lambda}( z_{1}, z_{2} , ... , z_{m} ) = \frac{  a_{\lambda}(z_{1}, z_{2} , ... , z_{m}) }{ \Delta(z_{1}, z_{2} , ... , z_{m}) }
	\end{align*}
	where $\Delta(z)$ is the Vandermode determinant.
	
	
	
\section{Unitary Quantum Double Calculation}\label{Unitary Quantum Double Calculation}
The matrix elements of $\hat{T}_{ij}$ are given by
	\begin{align*}
		\langle V_{1} , V_{2} |  \hat{T}_{ij} | V_{1}' , V_{2}' \rangle = \mathbb{E}_{U}[  \langle V_{1} ,V_{2} | UV'_{1}, V'_{2}U^{\dagger} \rangle  ] =  \frac{1}{d^{2}} \mathbb{E}_{U}[ \text{Tr}[ V^{\dagger}_{1}UV'_{1} ] \text{Tr}[ V^{\dagger}_{2}V'_{2} U^{\dagger} ]    ]
	\end{align*}
	We can evaluate this expression in closed form. Using the algebraic identity,
	\begin{align*}
		\text{Tr}[A]\text{Tr}[B] = \text{Tr}[ A \otimes B ]
	\end{align*}
	we can expand the product of traces as
	\begin{align*}
		\text{Tr}[V^{\dagger}_{1}UV'_{1}] \text{Tr}[V^{\dagger}_{2}V'_{2}U^{\dagger}] = \text{Tr}[V^{\dagger}_{1}UV'_{1}] \text{Tr}[V^{\dagger}_{2}V'_{2}U^{\dagger}]  = \text{Tr}[ V^{\dagger}_{1}UV'_{1} \otimes V^{\dagger}_{2}V'_{2} U^{\dagger}  ]
	\end{align*}
	where we have introduced an additional `copy' of the Hilbert space. We then have that
	\begin{align*}
		\mathbb{E}_{U}[ ( V'_{1}V^{\dagger}_{1}U  ) \otimes ( V^{\dagger}_{2}V'_{2}U^{\dagger}  )   ]  =  (V'_{1}V^{\dagger}_{1} \otimes \mathbb{I} ) \mathbb{E}_{U}[ U \otimes U^{\dagger}   ] (\mathbb{I} \otimes V^{\dagger}_{2}V'_{2} ) 
	\end{align*}
	Now, note that
	\begin{align*}
		\forall V \in U(d), \quad [V \otimes V^{\dagger} ]\mathbb{E}_{U}[ U \otimes U^{\dagger}   ] = \mathbb{E}_{U}[ U \otimes U^{\dagger}   ] [V \otimes V^{\dagger} ]
	\end{align*}
	Thus, using a `twisted' variant of Schur-Weyl duality \ref{More_Schur-Weyl_Duality}, we have that
	\begin{align*}
		\mathbb{E}_{U}[ U \otimes U^{\dagger}   ] = \alpha \mathbb{I}_{d} + \beta \hat{P}
	\end{align*}
	where $\hat{P}$ is the swap and conjugate operator and $\alpha$ and $\beta$ are constants. We then have that,
	\begin{align*}
		\langle V_{1} , V_{2} |  \hat{T}_{ij} | V_{1}' , V_{2}' \rangle = \text{Tr}[ (V'_{1}V^{\dagger}_{1} \otimes \mathbb{I} )[ \alpha \mathbb{I}_{d} + \beta \hat{S} ](\mathbb{I} \otimes V^{\dagger}_{2}V'_{2} ) ]  = \alpha \text{Tr}[ V'_{1}V^{\dagger}_{1} \otimes V'_{2}V^{\dagger}_{2}  ] + \beta d \text{Tr}[ V'_{1}V^{\dagger}_{2}V'_{2}V^{\dagger}_{1}    ]
	\end{align*}
	
	\section{Multi-Linear Algebra}
	We briefly review some multi-linear algebra concepts and operations on tensor product spaces. We specifically discuss partial transpose and partial conjugation, which are some standard tools in quantum information theory \cite{Nielsen_2000_Quantum}.
	
	\subsection{Partial Trace}
	
	The partial trace is a standard tool in quantum information theory \cite{Nielsen_2000_Quantum}. Let $H = H_{A} \otimes H_{B}$ be a Hilbert space composed of the $H_{A}$ and $H_{B}$ Hilbert spaces. Let $O$ be an operator defined on $W$. The partial trace of an operator on the $H_{A}$ or $H_{B}$ subspace is then defined as
	\begin{align*}
		O^{(A)} = \text{Tr}_{B}[  O  ], \quad O^{(B)} = \text{Tr}_{A}[  O  ]
	\end{align*}
	respectively, where the matrix elements of the partially traced operators are defined as
	\begin{align*}
		O^{(A)}_{ij} = \sum_{k=1}^{d_{B}} O_{ik,jk}, \quad O^{(B)}_{ij} = \sum_{k=1}^{d_{A}} O_{ki,kj}
	\end{align*}
	An operator $O$ is said to be separable if $O = O_{A} \otimes O_{B}$ factorizes. Partial traces of separable operators satisfy
	\begin{align*}
		O^{(A)} = \text{Tr}_{B}[  O  ] = \text{Tr}[  O_{B}  ] O_{A} , \quad O^{(B)} = \text{Tr}_{A}[  O  ]= \text{Tr}[  O_{A}  ] O_{B}
	\end{align*}
	A generic operator is not separable. However, via the operator-Schmidt decomposition.
	
	\begin{theorem}[Operator Schmidt-Decomposition]\label{Operator Schmidt-Decomposition}
		Let $O$ be an operator defined on the $V \otimes V$ tensor product space. The operator $O$ can always be written as
		\begin{align*}
			O = \sum_{\ell=1}^{ N_{O} } p_{\ell} A_{\ell} \otimes B_{\ell}
		\end{align*}
		where $p_{\ell}$ are positive real numbers and the operators $A_{\ell}$ and $B_{\ell}$ are orthogonal on the $V$ subspaces,
		\begin{align*}
			\text{Tr}[ A^{\dagger}_{\ell}A_{\ell'} ] = \delta_{\ell\ell'} = \text{Tr}[ B^{\dagger}_{\ell}B_{\ell'} ] 
		\end{align*}
		the integer $N_{O}$ (the rank of the matrix) counts the minimum number of tensor product operators needed to decompose $O$. $N_{O}$ is called the Schmidt number of the operator $O$.
	\end{theorem}
	\noindent
	The partial trace operation satisfies a uniqueness property. 
	\begin{theorem}[ Uniqueness of Partial Trace ]\label{Uniqueness_Partial_Trace}
		The partial trace is the unique linear map 
		\begin{align}
			\text{Tr}_{B}: L( A \otimes B ) \rightarrow L(A)
		\end{align}
		that satisfies the property
		\begin{align*}
			\forall H_{B} \in L(B), \enspace \forall H_{A} \in L(A), \quad \text{Tr}_{B}[  H_{A} \otimes H_{B}  ] = \text{Tr}[ H_{B} ] H_{A}
		\end{align*}
	\end{theorem}

	\subsection{Tensor Permutation Operators and Symmetric Group Representations }
	
	Let $W = V^{\otimes k}$ be a vector space that is the $k$-fold tensor product of $V$. For each permutation $\sigma \in S_{k}$, we define the operator $\hat{S}_{\sigma}$ with action on the tensor product basis via permutation
	\begin{align*}
		\forall \sigma \in S_{k}, \quad \hat{S}_{\sigma} | i_{1} i_{2} ... i_{k} \rangle =  | i_{\sigma(1)} i_{\sigma(2)}... i_{\sigma(k)}\rangle
	\end{align*}
	The operators $\hat{S}_{\sigma}$ form a unitary reducible representation of the group $S_{n}$. Specifically, the permutation representation will decompose as
	\begin{align*}
		( \hat{S}_{\sigma} , V^{\otimes k} ) \cong \bigoplus_{ \lambda \vdash k } c^{\lambda}_{(k,d) } \lambda
	\end{align*}
	with $ c^{\lambda}_{(k,d) } $ counting the muplicity of the irreducible $\lambda$ representation in $( \hat{S}_{\sigma} , V^{\otimes k} ) $. The character of the $( \hat{S}_{\sigma} , V^{\otimes k} )$ representation is given by
	\begin{align*}
		\chi( \sigma ) = \text{Tr}[ \hat{S}_{\sigma} ] = d^{ f(\sigma) } 
	\end{align*}
	where $f(\sigma)$ is the number of fixed points of the permutation $\sigma$. Thus,
	\begin{align*}
		c^{\lambda}_{ (k,d) } = \sum_{ \sigma \in S_{k} } \chi_{\lambda}( \sigma )d^{ f(\sigma) }
	\end{align*}
	where $\chi_{\lambda}( \sigma ): S_{k} \rightarrow \mathbb{C}$ is the character of the $\lambda$ irreducible.
	
	
	\subsubsection{$N=2$ Case}
	For $N=2$, $S_{2} \cong \mathbb{Z}_{2} $ is isomorphic to the cyclic group of order two. There are two permutation operators, $\mathbb{1}$ and $\hat{S}$. The operator $\hat{S}$ permutes tensor product indices with $\hat{S}| ij \rangle = | ji \rangle$. Note that
	\begin{align*}
		\hat{S}^{2} = \mathbb{1}
	\end{align*}
	Thus, $\hat{S}$ has eigenvalues $\pm 1$.
	All representations of $S_{2}$ are one dimensional. There are two irreducible representations, the trivial and sign representation. The tensor product space then decomposes as
	\begin{align*}
		V \otimes V = [ \frac{d(d+1)}{2} V_{+} ]\bigoplus [ \frac{d(d-1)}{2} ] V_{-}
	\end{align*}
	so that the tensor permutation space decomposes into $\frac{d(d+1)}{2}$ copies of the symmetric space and $\frac{d(d-1)}{2}$ copies of the anti-symmetric space. The projection operators into the $V_{+}$ and $V_{-}$ subspaces are given by
	\begin{align*}
		\hat{S}_{+} = \frac{1}{\sqrt{2}}( \mathbb{1}_{d \times d} + \hat{S} ) \quad 	\hat{S}_{-} = \frac{1}{\sqrt{2}}( \mathbb{1}_{d \times d} - \hat{S} ) 
	\end{align*}
	respectively. The projection operators are normalized to satisfy the relations $\hat{S}_{\pm}^{2} = \hat{S}_{\pm}$. Using Young diagrams, the irreducible representations are representation as the partitions $\lambda \vdash 2$, as shown in \ref{irriducibles of s2}.
	
	\begin{figure}[h]
		\centering
		$V_{+} \cong V_{ (2) } \cong $
		\begin{ytableau}
			1  & 2\\ 
		\end{ytableau}, \quad $V_{-} \cong V_{ (1,1) } \cong $
		\begin{ytableau}
			1 \\ 
			2 \\
		\end{ytableau}
		\caption{ Irreducible Representations of $S_{2}$ and corresponding Young Diagrams  }\label{irriducibles of s2}
	\end{figure}
	
	\subsection{Tensor Product Rules}
	\noindent
	The tensor product rules for the group $S_{2}$ are trivial. Using characters, we have that
	\begin{align*}
		V_{+} \otimes V_{+} = V_{+}, \quad V_{+} \otimes V_{-} = V_{-}, \quad V_{-} \otimes V_{-} = V_{+} 
	\end{align*}
	so that $C^{++}_{+} = 1, \enspace C^{+-}_{-} = C^{+-}_{-} = 1, \enspace C^{--}_{+}= 1$ and all other tensor product multiplicities are zero.
	
	\subsection{ Computing Branching and Induction Rules of $S_{2} \times S_{2} \subseteq S_{4}$ }\label{Rules for bipartite}
	There are five irreducible representations of $S_{4}$.
	\begin{figure}
		\centering
		\begin{ytableau}
			1 & 2 & 3 & 4 \\ 
		\end{ytableau} $\rightarrow$ \begin{ytableau}
			1 & 2 \\ 
		\end{ytableau} $\otimes$ \begin{ytableau}
			1 & 2\\ 
		\end{ytableau},
		\quad \quad 
		\begin{ytableau}
			1 \\ 
			2 \\
			3 \\
			4 \\
		\end{ytableau} $\rightarrow$ \begin{ytableau}
			1 \\ 
			2 \\
		\end{ytableau} $\otimes$ \begin{ytableau}
			1 \\ 
			2 \\
		\end{ytableau}, \quad \quad \quad
		\begin{ytableau}
			1 & 2 \\ 
			3 & 4\\
		\end{ytableau} $\rightarrow$  (
		\begin{ytableau}
			1 & 2 \\
		\end{ytableau} $\otimes$ \begin{ytableau}
			1 & 2 \\
		\end{ytableau} )
		$\oplus$
		(
		\begin{ytableau}
			1 \\ 
			2 \\
		\end{ytableau} $\otimes$ \begin{ytableau}
			1 \\ 
			2 \\
		\end{ytableau} ) 
		\hphantom{} \newline
		\hphantom{} \newline
		\begin{ytableau}
			1 & 2\\ 
			3 \\
			4 \\
		\end{ytableau} $\rightarrow$ ( \begin{ytableau}
			1 & 2 \\
		\end{ytableau} $\otimes$ \begin{ytableau}
			1 & 2 \\
		\end{ytableau}  ) $\oplus$ (
		\begin{ytableau}
			1 & 2 \\
		\end{ytableau}  $\otimes$ \begin{ytableau}
			1 \\ 
			2 \\
		\end{ytableau} )  $\oplus$ (
		\begin{ytableau}
			1 \\ 
			2 \\
		\end{ytableau} $\otimes$ \begin{ytableau}
			1 &2 \\ 
		\end{ytableau} ), \\ 
		\hphantom{} \newline
		\hphantom{} \newline
		\begin{ytableau}
			1 & 2 & 3  \\ 
			4  \\
		\end{ytableau} $\rightarrow$ (
		\begin{ytableau}
			1 \\
			2 \\
		\end{ytableau} $\otimes$ \begin{ytableau}
			1 \\
			2 \\
		\end{ytableau} ) $\oplus$ (
		\begin{ytableau}
			1 \\ 
			2 \\
		\end{ytableau} $\otimes$ \begin{ytableau}
			1 & 2 \\
		\end{ytableau} ) $\oplus$ (
		\begin{ytableau}
			1 & 2 \\
		\end{ytableau} $\otimes$ \begin{ytableau}
			1 \\ 
			2 \\
		\end{ytableau} )
		\caption{ Under the group restriction operation of $S_{4}$ to $S_{2} \times S_{2}$, The five irreducible representations $\lambda \vdash 4$ of $S_{4}$ decompose into direct sums of tensor products of $S_{2}$ irreducible representations.  }\label{Sfourestriction}
	\end{figure}
	The character table of irriducibles of $S_{4}$. The group $S_{4}$ has five conjugacy classes.
	
	\begin{table}
		\centering
		\begin{tabular}{ |p{4.23cm}|c|c|c|c|c|  }
			\hline
			\multicolumn{6}{|c|}{ \textbf{ Character Table of Irreducible Representations of $S_{4}$ }  } \\
			\hline
			Character & $e$,(size=1) & $(12)$,(size=6) & $(12)(34)$,(size=3) & $(123)$,(size=8) & $(1234)$,(size=6) \\
			\hline
			\hline
			$\chi_{(4)}$ & $ 1 $ & $1$ & $1$ & $1$ & $1$  \\
			\hline
			$\chi_{(1,1,1,1)}$ & $1$ & $-1$ & $1$ & $-1$ & $1$  \\
			\hline
			$\chi_{(2,2)}$ & $ 2 $ & $ 0 $ & $2$ & $-1$ & $0$  \\
			\hline
			$\chi_{(2,1,1)}$ & $3$ & $1$ & $-1$ & $0$ & $-1$  \\
			\hline
			$\chi_{(3,1)}$ & $3$ & $-1$ & $-1$ & $0$ & $1$  \\
			\hline
		\end{tabular}
		\caption{ Character Table of $S_{4}$ for irreducible representations $\lambda \vdash 4$.   }
	\end{table}
	\noindent
	Evaluated on the $S_{2} \times S_{2}$ subgroup, we have that
	\begin{align*}
		& \chi_{(4)}[ (e) (e) ] = 1, \quad \chi_{(4)}[ (12) (e) ] = 1, \quad \chi_{(4)}[ (e) (34) ] = 1, \quad \chi_{(4)}[ (12) (34) ] = 1 \\
		& \chi_{(1,1,1,1)}[ (e) (e) ] = 1, \quad \chi_{(1,1,1,1)}[ (12) (e) ] = -1, \quad \chi_{(1,1,1,1)}[ (e) (34) ] = -1, \quad \chi_{(1,1,1,1)}[ (12) (34) ] = 1 \\
		& \chi_{ (2,2) }[ (e)(e) ] = 2, \quad \chi_{ (2,2) }[ (12)(e) ] = 0, \quad \chi_{ (2,2) }[ (e)(12) ] = 0, \quad \chi_{ (2,2) }[ (12)(12) ] = 2 \\
		& \chi_{(2,1,1)}[ (e) (e) ] = 3, \quad \chi_{(2,1,1)}[ (12) (e) ] = 1, \quad \chi_{(2,1,1)}[ (e) (12) ] = 1, \quad \chi_{(2,1,1)}[ (12) (12) ] = -1, \\
		& \chi_{(3,1)}[ (e) (e) ] = 3, \quad \chi_{(3,1)}[ (12) (e) ] = -1, \quad \chi_{(3,1)}[ (e) (12) ] = -1, \quad \chi_{(3,1)}[ (12) (12) ] = -1,
	\end{align*}
	Upon restriction to the subgroup $S_{2} \times S_{2} $ we have the following decomposition of $S_{4}$ irreducible representations,
	\begin{align*}
		& V_{(4)} \rightarrow V_{+} \otimes V_{+}, \quad V_{(1,1,1,1)} \rightarrow V_{-} \otimes V_{-}, ,\quad V_{(2,2)} \rightarrow ( V_{+} \otimes V_{+} ) \oplus ( V_{-} \otimes V_{-} ) \\
		& V_{(2,1,1)} \rightarrow  ( V_{+} \otimes V_{+} ) \oplus ( V_{+} \otimes V_{-} ) \oplus ( V_{-} \otimes V_{+} ), \quad V_{(3,1)} \rightarrow ( V_{-} \otimes V_{-} ) \oplus ( V_{+} \otimes V_{-} ) \oplus ( V_{-} \otimes V_{+} ) \\
	\end{align*}
	This is shown diagrammatically in \ref{Sfourestriction}. Thus, the only non-zero branching rules are given by
	\begin{align*}
		& B^{ ++ }_{(4)} = 1, \quad B^{--}_{ (1,1,1,1) } = 1,\ \quad B^{++}_{ (2,2) } = B^{--}_{ (2,2) } = 1  \\
		& B^{--}_{ (3,1) } = B^{+-}_{ (3,1) } = B^{-+}_{ (3,1) } = 1, \quad B^{++}_{ (2,1,1,1) } =  B^{+-}_{ (2,1,1) } =  B^{-+}_{ (2,1,1) } = 1 \\
	\end{align*}

\bibliographystyle{plain} 
\bibliography{refs}

\begin{thebibliography}{10}

\bibitem{Aalsma2021Shocks}
L.~Aalsma, A.~Cole, E.~Morvan, J.~P. van~der Schaar, and G.~Shiu.
\newblock Shocks and information exchange in de sitter space.
\newblock {\em Journal of High Energy Physics}, 2021(10), October 2021.

\bibitem{Advani_2017}
Madhu~S. Advani and Andrew~M. Saxe.
\newblock High-dimensional dynamics of generalization error in neural networks,
  2017.

\bibitem{Ag_n_2021}
Cesar~A. Agón and Márk Mezei.
\newblock Bit threads and the membrane theory of entanglement dynamics.
\newblock {\em Journal of High Energy Physics}, 2021(11), November 2021.

\bibitem{Alexander2023Makes}
Yotam Alexander, Nimrod De~La Vega, Noam Razin, and Nadav Cohen.
\newblock What makes data suitable for a locally connected neural network? a
  necessary and sufficient condition based on quantum entanglement, 2023.

\bibitem{Altland1997Nonstandard}
Alexander Altland and Martin~R. Zirnbauer.
\newblock Nonstandard symmetry classes in mesoscopic normal-superconducting
  hybrid structures.
\newblock {\em Phys. Rev. B}, 55:1142--1161, Jan 1997.

\bibitem{baskerville2023random}
Nicholas~P Baskerville.
\newblock Random matrix theory and the loss surfaces of neural networks, 2023.

\bibitem{Baskerville_2022}
Nicholas~P. Baskerville, Diego Granziol, and Jonathan~P. Keating.
\newblock Appearance of random matrix theory in deep learning.
\newblock {\em Physica A: Statistical Mechanics and its Applications},
  590:126742, mar 2022.

\bibitem{Baskerville_2022_2}
Nicholas~P Baskerville, Jonathan~P Keating, Francesco Mezzadri, Joseph
  Najnudel, and Diego Granziol.
\newblock Universal characteristics of deep neural network loss surfaces from
  random matrix theory, 2022.

\bibitem{Beenakker1997Randomquantumtransport}
C.~W.~J. Beenakker.
\newblock Random-matrix theory of quantum transport.
\newblock {\em Rev. Mod. Phys.}, 69:731--808, Jul 1997.

\bibitem{Berry1999Hequalsxp}
M.~V. Berry and J.~P. Keating.
\newblock {\em H=xp and the Riemann Zeros}, pages 355--367.
\newblock Springer US, Boston, MA, 1999.

\bibitem{Berry_2023_Riemann}
M.~V. Berry and J.~P. Keating.
\newblock The riemann zeros and eigenvalue asymptotics.
\newblock {\em SIAM Review}, 41(2):236--266, 1999.

\bibitem{Blake2023Page}
Mike Blake and Anthony~P. Thompson.
\newblock The page curve from the entanglement membrane, 2023.

\bibitem{bouchaud2009financial}
J.~P. Bouchaud and M.~Potters.
\newblock Financial applications of random matrix theory: a short review, 2009.

\bibitem{Byrnes_2022}
Niall Byrnes and Matthew~R. Foreman.
\newblock Random matrix theory of polarized light scattering in disordered
  media.
\newblock {\em Waves in Random and Complex Media}, pages 1--29, dec 2022.

\bibitem{Ceccherini_2008_Harmonic}
T.~Ceccherini-Silberstein, F.~Scarabotti, and F.~Tolli.
\newblock {\em Harmonic Analysis on Finite Groups: Representation Theory,
  Gelfand Pairs and Markov Chains}.
\newblock Cambridge Studies in Advanced Mathematics. Cambridge University
  Press, 2008.

\bibitem{Chen_2022}
Liyuan Chen, Roy~J. Garcia, Kaifeng Bu, and Arthur Jaffe.
\newblock Magic of random matrix product states, 2022.

\bibitem{Collins_2016}
Benoit Collins and Ion Nechita.
\newblock Random matrix techniques in quantum information theory.
\newblock {\em Journal of Mathematical Physics}, 57(1):015215, jan 2016.

\bibitem{Collins_2023_II}
Benoît Collins, Razvan Gurau, and Luca Lionni.
\newblock The tensor harish-chandra–itzykson–zuber integral ii: Detecting
  entanglement in large quantum systems.
\newblock {\em Communications in Mathematical Physics}, 401(1):669–716,
  February 2023.

\bibitem{Collins_2023}
Benoît Collins, Razvan~G. Gurau, and Luca Lionni.
\newblock The tensor harish-chandra–itzykson–zuber integral i: Weingarten
  calculus and a generalization of monotone hurwitz numbers.
\newblock {\em Journal of the European Mathematical Society}, January 2023.

\bibitem{Comets_2019}
Francis Comets, Giambattista Giacomin, and Rafael~L. Greenblatt.
\newblock Continuum limit of random matrix products in statistical mechanics of
  disordered systems.
\newblock {\em Communications in Mathematical Physics}, 369(1):171--219, may
  2019.

\bibitem{Conrey2006Random}
J.~Brian Conrey, Jon~P. Keating, Michael~O. Rubinstein, and Nina~C. Snaith.
\newblock Random matrix theory and the fourier coefficients of half-integral
  weight forms, 2006.

\bibitem{Cottrell_2019theromfield}
William Cottrell, Ben Freivogel, Diego~M. Hofman, and Sagar~F. Lokhande.
\newblock How to build the thermofield double state.
\newblock {\em Journal of High Energy Physics}, 2019(2), February 2019.

\bibitem{DiFrancesco1997conformal}
Philippe Di~Francesco, Pierre Mathieu, and David Sénéchal.
\newblock {\em {Conformal field theory}}.
\newblock Graduate texts in contemporary physics. Springer, New York, NY, 1997.

\bibitem{Dohmatob_2022}
Elvis Dohmatob and Alberto Bietti.
\newblock On the (non-)robustness of two-layer neural networks in different
  learning regimes, 2022.

\bibitem{Dyson_2004_Threefold}
Freeman~J. Dyson.
\newblock {The Threefold Way. Algebraic Structure of Symmetry Groups and
  Ensembles in Quantum Mechanics}.
\newblock {\em Journal of Mathematical Physics}, 3(6):1199--1215, 12 2004.

\bibitem{Efetov_2005}
K.~B. Efetov.
\newblock Random matrices and supersymmetry in disordered systems.
\newblock 2005.

\bibitem{Ergun2009spectra}
G.~Ergun and R.~Kuehn.
\newblock Spectra of modular random graphs, 2009.

\bibitem{Fava_2023_designs}
Michele Fava, Jorge Kurchan, and Silvia Pappalardi.
\newblock Designs via free probability, 2023.

\bibitem{Feinberg_2021}
Joshua Feinberg and Roman Riser.
\newblock Dynamics of disordered mechanical systems with large connectivity,
  free probability theory, and quasi-hermitian random matrices.
\newblock {\em Annals of Physics}, 435:168456, dec 2021.

\bibitem{Gao2021Traversable}
Ping Gao and Daniel~Louis Jafferis.
\newblock A traversable wormhole teleportation protocol in the syk model, 2021.

\bibitem{Gao_2017_Traversable}
Ping Gao, Daniel~Louis Jafferis, and Aron~C. Wall.
\newblock Traversable wormholes via a double trace deformation.
\newblock {\em Journal of High Energy Physics}, 2017(12), December 2017.

\bibitem{Gharibyan_2018}
Hrant Gharibyan, Masanori Hanada, Stephen~H. Shenker, and Masaki Tezuka.
\newblock Onset of random matrix behavior in scrambling systems.
\newblock {\em Journal of High Energy Physics}, 2018(7), July 2018.

\bibitem{Goldstone1962Broken}
Jeffrey Goldstone, Abdus Salam, and Steven Weinberg.
\newblock Broken symmetries.
\newblock {\em Phys. Rev.}, 127:965--970, Aug 1962.

\bibitem{Gong_2022}
Zongping Gong, Adam Nahum, and Lorenzo Piroli.
\newblock Coarse-grained entanglement and operator growth in anomalous
  dynamics.
\newblock {\em Physical Review Letters}, 128(8), February 2022.

\bibitem{Guhr1998RMTquantum}
Thomas Guhr, Axel Müller–Groeling, and Hans~A. Weidenmüller.
\newblock Random-matrix theories in quantum physics: common concepts.
\newblock {\em Physics Reports}, 299(4–6):189–425, June 1998.

\bibitem{Hall_2015_Lie}
B.~Hall.
\newblock {\em Lie Groups, Lie Algebras, and Representations: An Elementary
  Introduction}.
\newblock Graduate Texts in Mathematics. Springer International Publishing,
  2015.

\bibitem{hartwell1999molecular}
Leland~H. Hartwell, John~J. Hopfield, Stanislas Leibler, and Andrew~W. Murray.
\newblock From molecular to modular cell biology.
\newblock {\em Nature}, 402:C47--C52, 1999.

\bibitem{Itzykson1979Planar}
C.~Itzykson and J.~B. Zuber.
\newblock {The Planar Approximation. 2.}
\newblock {\em J. Math. Phys.}, 21:411, 1980.

\bibitem{Jafferis_2022_Traversable}
Daniel Jafferis, Alexander Zlokapa, Joseph Lykken, David Kolchmeyer, Samantha
  Davis, Nikolai Lauk, Hartmut Neven, and Maria Spiropulu.
\newblock Traversable wormhole dynamics on a quantum processor.
\newblock {\em Nature}, 612:51--55, 11 2022.

\bibitem{Jafferis2023jt_gravity}
Daniel~Louis Jafferis, David~K. Kolchmeyer, Baur Mukhametzhanov, and Julian
  Sonner.
\newblock Jt gravity with matter, generalized eth, and random matrices, 2023.

\bibitem{Jafferis2023MatrixModels}
Daniel~Louis Jafferis, David~K. Kolchmeyer, Baur Mukhametzhanov, and Julian
  Sonner.
\newblock Matrix models for eigenstate thermalization.
\newblock {\em Phys. Rev. X}, 13:031033, Sep 2023.

\bibitem{james1981representation}
G.D. James and A.~Kerber.
\newblock {\em The Representation Theory of the Symmetric Group}.
\newblock Encyclopedia of mathematics and its applications. Addison-Wesley
  Publishing Company, Advanced Book Program, 1981.

\bibitem{Jaynes_1982}
E.T. Jaynes.
\newblock On the rationale of maximum-entropy methods.
\newblock {\em Proceedings of the IEEE}, 70(9):939--952, 1982.

\bibitem{Kanzieper_1998}
E.~Kanzieper and V.~Freilikher.
\newblock Two-band random matrices.
\newblock {\em Physical Review E}, 57(6):6604--6611, jun 1998.

\bibitem{Keating2006number}
J.~P. Keating.
\newblock Random matrices and number theory.
\newblock In {\'E}douard Br{\'e}zin, Vladimir Kazakov, Didina Serban, Paul
  Wiegmann, and Anton Zabrodin, editors, {\em Applications of Random Matrices
  in Physics}, pages 1--32, Dordrecht, 2006. Springer Netherlands.

\bibitem{Kitaev_2003Fault}
A.Yu. Kitaev.
\newblock Fault-tolerant quantum computation by anyons.
\newblock {\em Annals of Physics}, 303(1):2–30, January 2003.

\bibitem{Krblek2000mexico}
Milan Krbálek and Petr Seba.
\newblock The statistical properties of the city transport in cuernavaca
  (mexico) and random matrix ensembles.
\newblock {\em Journal of Physics A: Mathematical and General},
  33(26):L229–L234, June 2000.

\bibitem{Kuhn2011Spectra}
Reimer Kühn and Jort van Mourik.
\newblock Spectra of modular and small-world matrices.
\newblock {\em Journal of Physics A: Mathematical and Theoretical},
  44(16):165205, March 2011.

\bibitem{Lahtinen_2017}
Ville Lahtinen and Jiannis Pachos.
\newblock A short introduction to topological quantum computation.
\newblock {\em SciPost Physics}, 3(3), September 2017.

\bibitem{Landau_1991_Quantum}
L.D. Landau and Lifshitz.
\newblock {\em Quantum Mechanics: Non-Relativistic Theory}.
\newblock Course of theoretical physics. Elsevier Science, 1991.

\bibitem{Li_2022}
Jimin~L. Li, Dominic~C. Rose, Juan~P. Garrahan, and David~J. Luitz.
\newblock Random matrix theory for quantum and classical metastability in local
  liouvillians.
\newblock {\em Physical Review B}, 105(18), may 2022.

\bibitem{Louart_2017}
Cosme Louart, Zhenyu Liao, and Romain Couillet.
\newblock A random matrix approach to neural networks, 2017.

\bibitem{Maldacena1999LargeN}
Juan Maldacena.
\newblock {\em International Journal of Theoretical Physics},
  38(4):1113–1133, 1999.

\bibitem{Mann1989gh}
Robert~B. Mann, A.~Shiekh, and L.~Tarasov.
\newblock {Classical and Quantum Properties of Two-dimensional Black Holes}.
\newblock {\em Nucl. Phys. B}, 341:134--154, 1990.

\bibitem{mardia2009directional}
K.V. Mardia and P.E. Jupp.
\newblock {\em Directional Statistics}.
\newblock Wiley Series in Probability and Statistics. Wiley, 2009.

\bibitem{Martin_2021_Implict}
Charles~H. Martin and Michael~W. Mahoney.
\newblock Implicit self-regularization in deep neural networks: Evidence from
  random matrix theory and implications for learning.
\newblock {\em Journal of Machine Learning Research}, 22(165):1--73, 2021.

\bibitem{Mehta_2004}
M.L. Mehta.
\newblock {\em Random Matrices}.
\newblock ISSN. Elsevier Science, 2004.

\bibitem{Mezei_2018}
Márk Mezei.
\newblock Membrane theory of entanglement dynamics from holography.
\newblock {\em Physical Review D}, 98(10), November 2018.

\bibitem{Mitchell_2005}
David~R. Mitchell, Christoph Adami, Waynn Lue, and Colin~P. Williams.
\newblock Random matrix model of adiabatic quantum computing.
\newblock {\em Physical Review A}, 71(5), may 2005.

\bibitem{Newman2006Modularity}
M.~E.~J. Newman.
\newblock Modularity and community structure in networks.
\newblock {\em Proceedings of the National Academy of Sciences},
  103(23):8577–8582, June 2006.

\bibitem{Nielsen_2000_Quantum}
Michael~A. Nielsen and Isaac~L. Chuang.
\newblock {\em Quantum Computation and Quantum Information}.
\newblock Cambridge University Press, 2000.

\bibitem{pachos2012introduction}
J.K. Pachos.
\newblock {\em Introduction to Topological Quantum Computation}.
\newblock Introduction to Topological Quantum Computation. Cambridge University
  Press, 2012.

\bibitem{Pan_2022}
Haining Pan, Jay~Deep Sau, and Sankar Das~Sarma.
\newblock Random matrix theory for the robustness, quantization, and end-to-end
  correlation of zero-bias conductance peaks in a class d ensemble.
\newblock {\em Phys. Rev. B}, 106:115413, Sep 2022.

\bibitem{Pappalardi_2022_Eigenstate}
Silvia Pappalardi, Laura Foini, and Jorge Kurchan.
\newblock Eigenstate thermalization hypothesis and free probability.
\newblock {\em Physical Review Letters}, 129(17), October 2022.

\bibitem{Porter1960NuclearSpectra}
C~E Porter and N~Rosenzweig.
\newblock Statistical properties of atomic and nuclear spectra.
\newblock {\em Ann. Acad. Sci. Fennicae. Ser. A VI}.

\bibitem{potters2005financial}
M.~Potters, J.~P. Bouchaud, and L.~Laloux.
\newblock Financial applications of random matrix theory: Old laces and new
  pieces, 2005.

\bibitem{Rampp2023Entanglement}
Michael~A. Rampp, Suhail~A. Rather, and Pieter~W. Claeys.
\newblock The entanglement membrane in exactly solvable lattice models, 2023.

\bibitem{Roberts_2017_Chaos}
Daniel~A. Roberts and Beni Yoshida.
\newblock Chaos and complexity by design.
\newblock {\em Journal of High Energy Physics}, 2017(4), April 2017.

\bibitem{Rudnick1996ZerosOP}
Ze{\'e}v Rudnick and Peter Sarnak.
\newblock Zeros of principal \$l\$-functions and random matrix theory.
\newblock {\em Duke Mathematical Journal}, 81:269--322, 1996.

\bibitem{saad2019jt}
Phil Saad, Stephen~H. Shenker, and Douglas Stanford.
\newblock Jt gravity as a matrix integral, 2019.

\bibitem{Stockmann1999Quantum}
H.J. Sckmann.
\newblock {\em Quantum Chaos: An Introduction}.
\newblock Cambridge nonlinear science series. Cambridge University Press, 1999.

\bibitem{Shenker2014Black}
Stephen~H. Shenker and Douglas Stanford.
\newblock Black holes and the butterfly effect.
\newblock {\em Journal of High Energy Physics}, 2014(3), March 2014.

\bibitem{Sierant2023Entanglement}
Piotr Sierant, Marco Schir\`o, Maciej Lewenstein, and Xhek Turkeshi.
\newblock Entanglement growth and minimal membranes in ($d+1$) random unitary
  circuits.
\newblock {\em Phys. Rev. Lett.}, 131:230403, Dec 2023.

\bibitem{Sierra_2011}
Germán Sierra and Javier Rodríguez-Laguna.
\newblock <mml:math xmlns:mml="http://www.w3.org/1998/math/mathml"
  display="inline"><mml:mi>h</mml:mi><mml:mo>=</mml:mo><mml:mi>x</mml:mi><mml:mi>p</mml:mi></mml:math>model
  revisited and the riemann zeros.
\newblock {\em Physical Review Letters}, 106(20), May 2011.

\bibitem{Srednicki_1994Chaos}
Mark Srednicki.
\newblock Chaos and quantum thermalization.
\newblock {\em Physical Review E}, 50(2):888–901, August 1994.

\bibitem{stanford2020jt}
Douglas Stanford and Edward Witten.
\newblock Jt gravity and the ensembles of random matrix theory, 2020.

\bibitem{Stotland_2008}
Alexander Stotland, Rangga Budoyo, Tal Peer, Tsampikos Kottos, and Doron Cohen.
\newblock The mesoscopic conductance of disordered rings, its random matrix
  theory and the generalized variable range hopping picture.
\newblock {\em Journal of Physics A: Mathematical and Theoretical},
  41(26):262001, jun 2008.

\bibitem{tao2021}
Terence Tao.
\newblock {\em Topics in random matrix theory}.
\newblock 2011.

\bibitem{Thamm_2022}
Matthias Thamm, Max Staats, and Bernd Rosenow.
\newblock Random matrix analysis of deep neural network weight matrices.
\newblock {\em Physical Review E}, 106(5), nov 2022.

\bibitem{Turiaci2023n2JTsupergravity}
Gustavo~J. Turiaci and Edward Witten.
\newblock $n=2$ jt supergravity and matrix models, 2023.

\bibitem{van_Handel2017Structured}
Ramon van Handel.
\newblock {\em Structured Random Matrices}, page 107–156.
\newblock Springer New York, 2017.

\bibitem{Vleeshouwers_2023}
Ward~L Vleeshouwers and Vladimir Gritsev.
\newblock Unitary matrix integrals, symmetric polynomials, and long-range
  random walks.
\newblock {\em Journal of Physics A: Mathematical and Theoretical},
  56(18):185002, April 2023.

\bibitem{Wei_2022}
Alexander Wei, Wei Hu, and Jacob Steinhardt.
\newblock More than a toy: Random matrix models predict how real-world neural
  representations generalize, 2022.

\bibitem{Weyl_1966_Classical}
HERMANN WEYL.
\newblock {\em The Classical Groups: Their Invariants and Representations}.
\newblock Princeton University Press, 1966.

\bibitem{Wigner_1955}
Eugene~P. Wigner.
\newblock Characteristic vectors of bordered matrices with infinite dimensions.
\newblock {\em Annals of Mathematics}, 62(3):548--564, 1955.

\bibitem{Xu2020Sparse}
Shenglong Xu, Leonard Susskind, Yuan Su, and Brian Swingle.
\newblock A sparse model of quantum holography, 2020.

\bibitem{Yadin_2023Thermodynamics}
Benjamin Yadin, Benjamin Morris, and Kay Brandner.
\newblock Thermodynamics of permutation-invariant quantum many-body systems: A
  group-theoretical framework.
\newblock {\em Physical Review Research}, 5(3), July 2023.

\bibitem{Yang_2020}
Fan Yang, Hongyang~R. Zhang, Sen Wu, Weijie~J. Su, and Christopher Ré.
\newblock Analysis of information transfer from heterogeneous sources via
  precise high-dimensional asymptotics, 2020.

\bibitem{Zee_2016}
A.~Zee.
\newblock {\em Group Theory in a Nutshell for Physicists}.
\newblock In a Nutshell. Princeton University Press, 2016.

\bibitem{Zhou_2020Entanglment}
Tianci Zhou and Adam Nahum.
\newblock Entanglement membrane in chaotic many-body systems.
\newblock {\em Physical Review X}, 10(3), September 2020.

\end{thebibliography}

\end{document}